\documentclass[12pt]{article}

\usepackage{amsmath}
\usepackage{slashed}
\usepackage[dvips]{graphicx}

\newcommand{\nn}{\nonumber}
\newcommand{\be}{\begin{equation}}
\newcommand{\ee}{\end{equation}}
\newcommand{\ba}{\begin{eqnarray}}
\newcommand{\ea}{\end{eqnarray}}
\newcommand{\ci}[1]{\cite{#1}}
\def\={\,=\,}

\def\vd{{\bf \Delta}_\perp}

\def\als{\alpha_s}
\def\ale{\alpha_{\rm em}}
\def\mev{\,{\rm MeV}}
\def\gev{\,{\rm GeV}}

\newcommand{\sla}{\hspace*{-0.20cm}/}

\newcommand{\da}{{DA}}

\def\muR{\mu^2_R}     
\def\muF{\mu^2_F}
\def\muO{\mu^2_0}
\newcommand{\tw}{\textwidth}                          
\newcommand{\req}[1]{(\ref{#1})}

\def\eps{\epsilon}

\def\sh{\hat{s}}
\def\uh{\hat{u}}
\def\th{\hat{t}}

\def\taub{\bar{\tau}}

\def\phiDA{\phi}

\begin{document}
\thispagestyle{empty}
\begin{flushright}
WU-B 18-00\\
February, 17  2018\\[20mm]
\end{flushright}

\begin{center}

{\Large\bf Twist-3 contributions to wide-angle \\[0.3em]
photoproduction of pions} \\
\vskip 15mm
P.\ Kroll \\[1em]
{\small {\it Fachbereich Physik, Universit\"at Wuppertal, D-42097 Wuppertal,
Germany}}\\[1em]
K.\  Passek-Kumeri\v{c}ki\\[1em]
{\small {\it Theoretical Physics Division, Rudjer Bo\v{s}kovi\'{c} Institute, 
HR-10002 Zagreb, Croatia}}\\

\end{center}

\begin{abstract}
We investigate wide-angle $\pi^0$ photoproduction within the handbag approach 
to twist-3 accuracy. In contrast to earlier work both the 2-particle as well as the 
3-particle twist-3 contributions are taken into account. It is shown that both are 
needed for consistent results that respect gauge invariance and crossing properties.
The numerical studies reveal the dominance of the twist-3 contribution. With it fair 
agreement with the recent CLAS measurement of the $\pi^0$ cross section is obtained. 
We briefly comment also on wide-angle photoproduction of other pseudoscalar mesons.
\end{abstract}

\section{Introduction}
\label{sec:introduction}
Since 1996 there are a lot of activities on the field of hard exclusive processes in conjunction 
with handbag factorization.
A vast amount of data on such processes have been accumulated from HERMES, COMPASS, BaBar
and BELLE and from  experiments performed at Jefferson Lab and  HERA. Many theoretical
studies of these processes have been carried through within the framework of the handbag 
approach in which the process amplitudes factorize in hard, perturbatively calculable 
subprocesses and soft hadron matrix elements, parametrized as generalized parton distributions (GPDs).

Of particular importance for the present work is wide-angle Compton scattering (WACS). 
There are reasonable arguments \ci{rad98,DFJK1} that for large Mandelstam variables, $s$, $-t$ 
and $-u$, the Compton amplitudes can be represented as a product of amplitudes for the subprocess,
Compton scattering off quarks, and form factors that represent $1/x$-moments of GPDs.
Since the GPDs in question, namely $H$, $E$ and $\widetilde H$, are known from an analysis 
of the form factors of the nucleon \ci{DK13} one can compute the Compton form factors and 
subsequently the Compton cross section as well as other observables for this process. The results 
of this parameter-free prediction \ci{DK13} agrees quite well with experiment \ci{hallA} given 
that the Mandelstam variables achieved in current experiments are not large as compared to a 
typical hadronic scale of 
order $1\gev^2$. An analogous calculation of wide-angle photoproduction of mesons however fails
\ci{huang00}: the cross sections are underestimated by about two orders of magnitude. An 
attempt to improve this result has been presented in \ci{signatures}: under the assumption of a 
vanishing contribution from the $q\bar{q}g$ Fock component of the meson (frequently termed the 
Wandzura-Wilczek approximation) the 2-particle twist-3 meson distribution amplitudes (\da s) 
have been taken into account along with the helicity-flip or transversity GPDs 
\ci{ji-hoodbhoy,diehl01}. The analysis however revealed that the corresponding 2-particle twist-3 
contribution is zero. Thus, this attempt turned out to be unsuccessful.

A HERMES measurement \ci{hermes08} of the asymmetry in electroproduction of positively charged 
pions, obtained with a transversely polarized target, indicated a strong contribution
from transversely polarized virtual photons which in the generalized Bjorken regime of
large photon virtuality, $Q^2$, and large photon-proton center-of-mass energy but fixed Bjorken-$x$
and $-t\ll Q^2$, is in principal suppressed by $1/Q^2$ in the cross section as compared to the 
asymptotically leading contribution from longitudinally polarized photons \ci{collins}. In \ci{GK5,GK6} 
it has been shown that the HERMES result on the asymmetry can be understood by the just the same 
dynamical mechanism, namely the combination of transversity GPDs and twist-3 pion \da s, 
that failed in wide-angle photoproduction as we mentioned above. We stress that in pion electroproduction
the mechanism in question is probed at large $Q^2$ but $t\to 0$ in contrast to photoproduction
where $-t$ (and $-u$) are large but $Q^2\to 0$. The twist-3 contribution is large in the case of
pions  because it is proportional to a mass parameter, $\mu_\pi$, which is related to the chiral
condensate
\be
\mu_\pi \= \frac{m_\pi^2}{m_u+m_d}
\label{eq:mass-parameter}
\ee
by means of the divergence of the axial vector current. Here, $m_i$ are current quark masses and 
$m_\pi$ denotes the mass of the pion. This parameter is 
large, about $2\,\gev$ at the scale $2\,\gev$. The transverse cross section for pion 
electroproduction is parametrically suppressed by $\mu_\pi^2/Q^2$ as compared to the longitudinal 
cross section. For the accessible range of $Q^2$ in current experiments the suppression factor is 
of order unity. Predictions for the $\pi^0$ electroproduction cross sections given in \ci{GK6} (see
also \ci{liuti}), revealed a transverse cross section  that is much larger than the longitudinal one. 
This prediction has been confirmed by a recent measurement of the separated $\pi^0$ cross sections 
performed by the Jefferson Lab Hall A collaboration \ci{defurne}. The longitudinal cross section is 
found to be compatible with zero within the experimental errors. A preliminary COMPASS result 
\ci{compass} for the unseparated cross section at a much larger center-of-mass energy but approximately 
the same $Q^2$ is, in tendency, in agreement with the Hall A findings. Thus, the same situation appears 
in both hard $\pi^0$ electroproduction and wide-angle $\pi^0$ photoproduction -  a leading-twist 
analysis fails badly in comparison with experiment at presently available hard scales.        
 
In view of these experimental and theoretical results on hard exclusive pion electroproduction  
a resumption of the investigation on the wide-angle meson photoproduction seems to be appropriate
and this is the purpose of the present work. It differs from the earlier work \ci{huang00,signatures}
by the inclusion of the full, genuine twist-3 contribution, i.e.\ its 2-particle as well as its 
3-particle part. Both parts are related to each other by the equation of motion \ci{braun90} and both 
are required in order to accomplish gauge invariance and crossing properties. In Sect.\ 2 we 
recapitulate  the handbag approach to photoproduction of uncharged pions to twist-3 accuracy. In 
the next section, Sect.\ 3, we discuss the large $-t$ behavior of the relevant helicity flip and 
non-flip GPDs and the corresponding form factors. The subprocess amplitudes to twist-3 accuracy 
are discussed in Sect. 4 and, in Sect.\ 5, results for the cross section and spin-dependent 
observables for photoproduction of the $\pi^0$ are presented. It is also commented on 
photoproduction of other mesons. The paper is finished with the usual summary. In appendix A the 
2- and 3-particle twist-3 \da s are discussed in some detail. The separate 2- and 3-particle 
twist-3 subprocess amplitudes are presented in App.\ B.

\section{The handbag mechanism}
\label{sec:handbag}
The handbag mechanism for wide-angle photoproduction of uncharged pions, $\gamma p \to \pi^0 p$, where 
$p$ denotes a proton, has been developed in \ci{huang00,signatures}. For a 
better comprehension of the present work we are going to recapitulate the main results and arguments 
for factorization of the photoproduction amplitude in hard subprocesses and soft form factors.

Prerequisite is that the Mandelstam variables $s$, $-t$ and $-u$ are much larger than $\Lambda^2$ where 
$\Lambda$ is a typical hadronic scale of order $1\,\gev$. It is of advantage to work in a symmetrical 
frame which is a center-of-mass frame (c.m.s.) rotated in such a way that the momenta of the ingoing 
($p$) and outgoing ($p'$) nucleons have the same light-cone plus components
\be
p\=\Big[p^+, \frac{m^2-t/4}{2p^+}, -\frac12 \vd\Big]\,, \qquad
p'\=\Big[p^+, \frac{m^2-t/4}{2p^+}, \phantom{-}\frac12 \vd\Big]\,,  
\ee
where $m$ is the mass of the proton. In this frame the skewness, defined by
\be
\xi\=\frac{(p-p')^+}{(p+p')^+}\,,
\ee
is zero. We assume restricted parton virtualities $k_i^2<\Lambda^2$ and intrinsic transverse parton 
momenta, $k_{\perp i}$, defined with respect to their parent hadron's momentum, which satisfy the 
condition $k^2_{\perp i}/x_i<\Lambda^2$. Here, $x_i$ denotes the momentum fraction that parton 
$i$ carries. On these premises one can show \ci{huang00} that the subprocess Mandelstam variables
$\sh$ and $\uh$ coincide with the ones for the full process, photoproduction of pions, up to 
corrections~\footnote{
        Possible corrections due to the proton mass have been discussed in \ci{huang02}.}
of order $\Lambda^2/s$
\be
\th\=t\,, \quad \sh\=(k_j+q)^2\simeq (p+q)^2\=s\,, \quad 
\uh\=(k'_j-q)^2\simeq (p'-q)^2\=u\,,
\ee
where $k_j$ and $k_j'=k_j+q-q'$ denote the momenta of the active partons, i.e., the in and out partons 
to which the photon couples; $q$ and $q'$ are the momenta of the photon and meson, respectively. 
Thus, the active partons are approximately on-shell, move collinear with their parent hadrons and carry 
a momentum fraction close to unity, $x_j, x'_j\simeq 1$. As in deeply virtual exclusive scattering, the 
physical situation is that of a hard parton-level subprocess, $\gamma q_a\to \pi^0 q_a$, and a soft 
emission and reabsorption of quarks from the proton. Up to corrections of order $\Lambda/\sqrt{-t}$ the 
light-cone helicity amplitudes for wide-angle photoproduction are then given by a product of subprocess 
amplitudes, ${\cal H}$, and form factors
\ba
{\cal M}_{0+,\mu +}&=& \frac{e_0}{2}\, \sum_\lambda \Big[
              {\cal H}_{0\lambda,\mu\lambda}\, \left( R_V^{\pi^0}(t)
              +2\lambda  \, R_A^{\pi^0}(t) \right) \nn\\
        &&    - 2\lambda \, \frac{\sqrt{-t}}{2m} \,
{\cal H}_{0-\lambda,\mu\lambda}\,\bar{S}^{\pi^0}_T(t)\Big]\,, \nn\\
{\cal M}_{0-,\mu +} &=& \frac{e_0}{2}\, \sum_\lambda \Big[
                 \frac{\sqrt{-t}}{2m} \, {\cal H}_{0\lambda,\mu\lambda}\, R_T^{\pi^0}(t) \nn\\
           && -2\lambda \, \frac{t}{2m^2} \, {\cal H}_{0-\lambda,\mu\lambda}\,S^{\pi^0}_S(t)\Big]
                              + e_0 {\cal H}_{0-,\mu +}\,S^{\pi^0}_T(t)\,,
\label{eq:amplitudes}
\ea 
where $\mu$ denotes the helicity of the photon, $\lambda$ the helicity of the active quark and 
$e_0$ the positron charge. Note that for the sake of legibility helicities are labeled by 
their signs only. The amplitudes for helicity configurations other than quoted in \req{eq:amplitudes} 
follow from parity invariance
\be
{\cal M}_{0-\nu',-\mu-\nu} \= (-1)^{\nu-\nu'} {\cal M}_{0\nu',\mu\nu}
\ee
An analogous relation holds for the subprocess amplitudes ${\cal H}$. The soft form factors, 
$R_i^{\pi^0}$ and $S_i^{\pi^0}$, are specific to photoproduction of uncharged pions. They represent 
$1/x$-moments of GPDs at zero skewness, where $x=(k_j+k'_j)^+/(p+p')^+$ is the average momentum 
fraction the two active quarks carry. The form factors parametrize the soft physics that controls
the emission from and reabsorption of a quark by the proton. They will be discussed in some detail
in the next section. The representation 
\req{eq:amplitudes}, which requires the dominance of the plus components of the 
proton matrix elements, is a non-trivial feature given that, in contrast to deep inelastic
lepton-nucleon and deep virtual exclusive processes, not only the plus components of 
the proton momenta but also their minus and transverse components are large in this case 
\ci{DFJK1}. The generalization of \req{eq:amplitudes} to photoproduction of other pseudoscalar
mesons is straightforward \ci{signatures}.

\section{GPDs and form factors at large $-t$}
\label{sec:gpds}
The form factors for an active quark of flavor $a$ are defined by \ci{huang00,signatures}
\ba
R_V^a(t) &=& \int^1_{-1}\frac{dx}{x}\, {\rm sign}(x) H^a(x,t)\,, \quad
S_T^a(t) \= \int^1_{-1}\frac{dx}{x}\, {\rm sign}(x) H_T^a(x,t)\,,           \nn\\
R_A^a(t) &=& \int^1_{-1}\frac{dx}{x}\, \widetilde{H}^a(x,t)\,,  \hspace*{0.12\tw} 
S_S^a(t) \= \int^1_{-1}\frac{dx}{x}\, {\rm sign}(x) \widetilde{H}_T^a(x,t)\,,  \nn\\
R_T^a(t) &=& \int^1_{-1}\frac{dx}{x}\, {\rm sign}(x) E^a(x,t)\,,  \quad
S_V^a(t) \= \int^1_{-1}\frac{dx}{x}\, {\rm sign}(x) E_T^a(x,t)\,.  
\label{eq:form-factors}
\ea
It is also convenient to introduce the combination
\be
\bar{S}^a_T(t) \= 2S^a_S(t) + S_V^a(t)
\label{eq:barS}
\ee
associated with the GPD $\bar{E}_T=2\widetilde{H}_T+E_T$.
The functions $H^a$, $\widetilde{H}^a$ and $E^a$ are the familiar helicity non-flip
GPDs at zero skewness whereas $H_T^a$, $\widetilde{H}^a_T$ and $E_T^a$ denote the helicity flip
or transversity GPDs. The skewness variable is omitted in the GPDs for convenience. The GPDs 
$\tilde{E}^a$ and $\tilde{E}_T^a$ and their associated form factors decouple in the symmetrical frame. 
Note that $x$ runs from -1 to +1. As usual a parton with a negative momentum fraction  is
reinterpreted as an antiproton with a positive momentum fraction. One has 
\be
K^{\bar{a}}(x,t)\=-K^a(-x,t) \qquad (x>0)
\ee
for all GPDs, $K$, except for $\widetilde{H}$ for which the relation
\be
\widetilde{H}^{\bar{a}}(x,t)\=\widetilde{H}^a(-x,t)  \qquad (x>0)
\ee
holds. Thus, the flavor form factors in \req{eq:form-factors} can also be written as
($F_i=R_V, R_A, \ldots S_V$)
\be
  F_i^a(t)\=\int_0^1 \frac{dx}{x} \big( K_i^a(x,t) - K_i^{\bar{a}}(x,t) \big)\,.
\ee
One notices that quarks and antiquarks contribute with opposite sign to photoproduction of
pseudoscalar mesons, i.e., only valence quarks contribute. This is to be contrasted
with Compton scattering \ci{DFJK1} or photoproduction of vector mesons \ci{huang00}
where they contribute with the same sign. This feature reflects the charge-conjugation
properties of the GPDs.

The flavor form factors are to be combined in form factors specific to a given process.
Thus, for the process on which we focus our interest, $\pi^0$ photoproduction off protons, 
the relevant combination of the flavor form factors is
\be
F_i^{\pi^0}(t) \= \frac1{\sqrt{2}} \big[e_u F_i^u(t) - e_d F_i^d(t) \big]\,.
\label{eq:pi0-form-factors}
\ee  
where $e_a$ is the charge of a quark of flavor $a$ in units of the positron charge, $e_0$.

In \ci{DK13} the GPDs $H$ and $E$ for valence quarks have been extracted from the data
on the magnetic form factors of the proton and the neutron and from the ratio of electric
and magnetic form factors exploiting the sum rules for the form factors
with the help of a parametrization of the zero skewness GPDs
\be
K_i^a\=k_i^a(x) \exp{[tf_i^a(x)]}\,.
\label{eq:GPD-ansatz}
\ee
In \ci{DK13,DFJK4} it is advocated for the following parametrization of the profile function
\be
f_i^a(x)\=\big(B_i^a - \alpha_i'{}^a\ln{x}\big)(1-x)^3 + A_i^a x(1-x)^2
\,,
\label{eq:profile}
\ee
with $A_i$, $B_i$ and $\alpha_i$ being the parameters discussed below.
The forward limit of the GPD $H^a$ is given by the flavor-a parton density, $q^a(x)$. On the other
hand the forward limit of $E^a$ is not accessible in deep-inelastic scattering and is, therefore, 
to be determined in the form factor analysis, too. For the parameterization, \req{eq:GPD-ansatz} and 
\req{eq:profile} there is a strong $x - t$ correlation in the GPD as has been discussed in 
\ci{DK13,DFJK4}. The GPDs at small $x$ control the behavior of the associated flavor form factors at 
small $-t$ whereas large $x$ determine their large $-t$ behavior which is required for wide-angle 
photoproduction of mesons. As is obvious from \req{eq:profile} at small $x$ the first term of the 
profile function dominates while at large $x$ its second term is important. In analyses of deeply
virtual exclusive processes, as electroproduction of photons or meson, for which only data at 
rather small $-t$ are available, the so-called Regge-like profile function is frequently used.
This profile function is just the first term of \req{eq:profile} with the factor $(1-x)^3$ being 
dropped. Clearly, in view 
of the $x-t$ correlation, we can only learn about the GPDs at small $x$ from data on deeply virtual 
exclusive processes;  an extrapolation to large $x$ and large $-t$ is dangerous and may lead to
misleading results~\footnote{
    As shown in \ci{DFJK4} the Regge-like profile function leads to an infinitely large 
    distance between the active quark and the cluster of spectators, i.e., it leads to a violation
    of  confinement.}.   
The parameter $A$ of the second term in \req{eq:profile} cannot be fixed from small $-t$  data. 
Information on the large $-t$ (large $x$) behavior of the GPDs is for instance obtained from the 
electromagnetic form factors of the nucleon.

After these preliminaries we move on to the discussion of the actual choice of the form factors:
$R_V^{\pi^0}$ and $R_T^{\pi^0}$ are evaluated from the GPDs derived in \ci{DK13}. 
The GPD $\widetilde{H}$ is only known for $-t$ less than about $3\,\gev^2$ \ci{DK13} from the data 
on the axial form factor \ci{kitagaki}. Data at larger $-t$ are to be expected from the upcoming 
Fermilab MINERvA experiment. From data on the helicity correlations $A_{LL}$ and/or $K_{LL}$ in 
wide-angle Compton scattering we may also learn about the large $-t$ behavior of the GPD 
$\widetilde H$ \ci{kroll17}. Measurements of these helicity correlations are 
planned at Jefferson Lab. On the basis of the parametrization \req{eq:GPD-ansatz} and \req{eq:profile}
(with the unpolarized parton densities replaced by the polarized ones) and the results on 
$\widetilde{H}$ given in \ci{DK13} several examples of the large $-t$ behavior of $\widetilde H$ are
discussed in  \ci{kroll17}. For the numerical estimates of observables for wide-angle photoproduction 
of pseudoscalar mesons to be presented in Sect. \ref{sec:pi0-results} below we will use example \#1 
quoted in \ci{kroll17}. 

\begin{table*}[t]
\renewcommand{\arraystretch}{1.4} 
\begin{center}
\begin{tabular}{| c || c  c|  c  c|}
     & $H_T^u$ & $H^d_T$ & $\bar{E}_T^u$ &  $\bar{E}_T^d$ \\[0.2em]
\hline   
 $N$                   & 0.78 & -1.01 &  4.83   & 3.57  \\[0.2em]
 $\alpha'\,[\gev^{-2}]$ & 0.45 &  0.45 &  0.45   & 0.45  \\[0.2em]
 $B\,[\gev^{-2}]$       & 0 & 0 & 0.50 &  0.50 \\[0.2em]
\hline
\end{tabular}
\end{center}
\caption{Parameters of the GPDs $H_T$ and $\bar{E}_T$ taken from \ci{GK6}.} 
\label{tab:1}
\renewcommand{\arraystretch}{1.0}   
\end{table*} 

In \ci{GK5,GK6} hard pion electroproduction has been studied and the valence quark GPDs $H_T$ and 
$\bar{E}_T$ at small $-t$  extracted. These GPDs are also parametrized as in \req{eq:GPD-ansatz}
and \req{eq:profile}. Their forward limits read
\ba
h_T^a &=& N_H^a x^{1/2} (1-x) \big[q^a(x) + \Delta q^a(x)\big]\,, \nn\\
\bar{e}_T^{\,a} &=& N_E^a x^{-\alpha_{e_T}^a} (1-x)^{\beta_{e_T}^a}\,.
\ea
The particular parametrization of the forward limit of $H_T$ guarantees that the Soffer bound
is respected. For the numerical studies the parton densities are taken from \ci{abm11} and \ci{dssv09}. 
The parameters of the GPDs $H_T$ and $\bar{E}_T$ are quoted in Tab.\ \ref{tab:1}. In addition there
are the parameters of the forward limit of $\bar{e}_T$:
\be
\alpha_{e_T}^u\=\alpha_{e_T}^u\=0.3\,,\qquad \beta^u_{e_T}\=4\,,\qquad \beta^d_{e_T}\=5\,.
\ee
which are also taken from \ci{GK6}.

\begin{figure}[t]
\begin{center}
\includegraphics[width=0.59\tw]{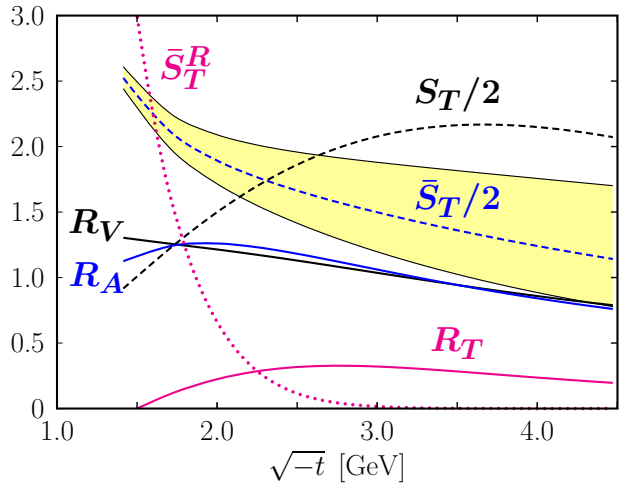}
\caption{\label{fig:FF} The form factors for $\pi^0$ photoproduction scaled by $t^2$. The 
dimension is $\gev^4$. For the transversity form factor the value of the parameter $A$ is $0.5\,\gev^{-2}$.
The upper (lower) edge of the band for $\bar{S}_T$ is evaluated from $A=0.3\, (0.7) \gev^{-2}$.}
\end{center}
\end{figure} 

\begin{table*}[t]
\renewcommand{\arraystretch}{1.4} 
\begin{center}
\begin{tabular}{| c || c  c  c  c  c|}
     & $R_V$ & $R_A$ &  $R_T$ &  $S_T$ & $\bar{S}_T$ \\[0.2em]
\hline   
 $u$ & 2.25  & 2.22 &  2.83    & 2.5 & 2.5   \\[0.2em]
 $d$ & 3.0   & 2.61 &  3.12    & 3.5 & 3.0    \\[0.2em]
\hline
\end{tabular}
\end{center}
\caption{The powers $d_i$ for the various form factors contributing
to the wide-angle photoproduction of pseudoscalar mesons.} 
\label{tab:2}
\renewcommand{\arraystretch}{1.0}   
\end{table*} 

As mentioned before and shown in Fig.\ \ref{fig:FF} the Regge-like profile function
leads to form factors that rapidly drop with $-t$. Clearly, also for wide-angle photoproduction of 
pseudoscalar mesons the second term of the profile function \req{eq:profile} is required. In the 
absence of any information on the parameter $A$ we tentatively choose for it the value 
$0.5\,\gev^{-2}$ in all cases. Fortunately, the dependence of the form factors $S_i$ on that 
parameter is rather mild in the range of $t$ relevant to current photoproduction experiments. This 
is demonstrated by the band for $\bar{S}_T^{\pi^0}$ evaluated from $A=0.3\,\gev^{-2}$ and 
$0.7\,\gev^{-2}$. The form factor $\bar{S}_T^{\pi^0}$ is rather large since $\bar{E}_T^u$ and 
$\bar{E}^d_T$ have the same sign and about the same normalization, a fact that is supported by 
results from lattice QCD \ci{QCDSF06}. This feature of $\bar{E}_T$ is also responsible for the 
dominance of this GPD in electroproduction of $\pi^0$,

With the help of the saddle point method \ci{DFJK4} one can show  that the moments of the GPDs, 
parametrized by \req{eq:GPD-ansatz} and \req{eq:profile}, behave power-law like:
\be
    F_i \sim 1/(-t)^{d_i}\,.
\ee
The power $d_i$ is determined by the power $\beta_i$ of the factor $1-x$ that characterizes the 
behavior of the forward limits of the GPDs for $x \to 1$ 
\be
d_i=(1+\beta_i)/2\,.
\ee
We stress that the power $\beta_i$ is fixed in a region of $x$ less than about 0.8. For larger $x$
there is no experimental information on the forward limits available at present. Therefore, the 
powers $\beta_i$ are to be considered rather as effective powers which are likely subject to 
change as soon as data at larger $x$ become available. The current powers $d_i$ are listed in 
Tab.\ \ref{tab:2}.

At present there is no information available on the GPD ${\widetilde H}_T$ and its associated from 
factor $S_S$. It has been neglected in the analysis of electroproduction of pseudoscalar mesons 
because its contribution is suppressed by a factor $t/(2m^2)$ (see also \req{eq:amplitudes}). 
However, this argument does no longer hold for wide-angle meson photoproduction since $-t$ is 
large. As an estimation of its significance we take 
$S^{\pi^0}_S\simeq\bar{S}^{\pi^0}_T/2$ ($S_V^{\pi^0}\simeq 0$, cf.\ \req{eq:barS}).

\section{The subprocess amplitudes}
\label{sec:subprocess}

\begin{figure}[t]
\begin{center}
\includegraphics[width=0.59\tw]{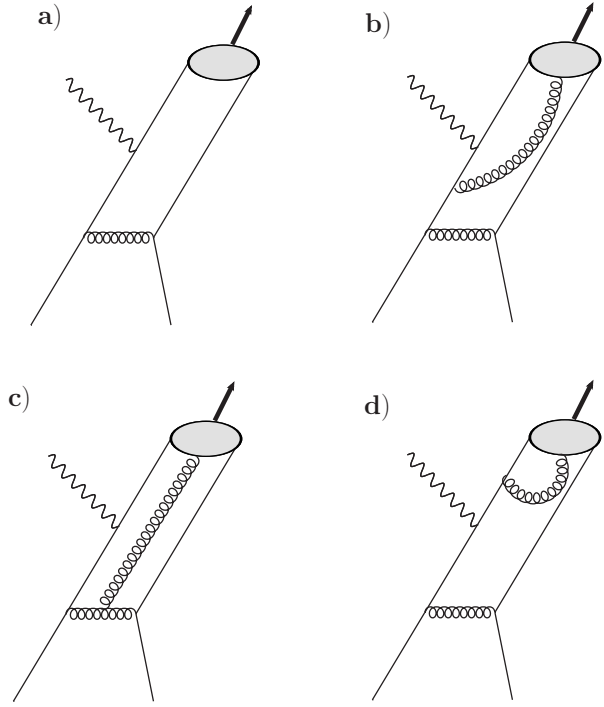}
\caption{\label{fig:graphs} Typical leading-order Feynman graphs for 
$\gamma q\to \pi^0 q$. a) for a 2-particle Fock component of a pseudoscalar meson.
b) and c) contribution from the $q\bar{q}g$ Fock component without and with 
triple gluon coupling. d) a soft contribution which is to be considered as
part of the 3-particle \da. }
\end{center}
\end{figure} 

We calculate the amplitudes for the subprocess $\gamma q_a\to \pi^0 q_a$ to twist-3 accuracy. 
In the definitions of the various vacuum-meson matrix elements as frequently done
in QCD calculations of exclusive processes, we are using light-cone (axial) gauge.
All possible Wilson lines become unity in that gauge. Our calculation method is similar to 
light-cone collinear factorization approach discussed in detail in \ci{wallon,anikin02} for the 
case of electroproduction of transversely polarized vector mesons. 

Typical lowest-order Feynman graphs for the process of interest are depicted in 
Fig.\ \ref{fig:graphs}. In particular the four graphs of type a) are relevant for the 2-particle 
contributions. 
With the help of the $q\bar{q}\to \pi^0$ projector  \ci{beneke,two-gluon}
\ba
{\cal P}_{2,fg}&=&\frac{f_\pi}{2\sqrt{2N_C}}\,\frac{\delta_{fg}}{\sqrt{N_C}}
                       \,\left\{\frac{\gamma_5}{\sqrt{2}}\,q\sla'\phiDA_\pi(\tau) 
                   + \mu_{\pi}\frac{\gamma_5}{\sqrt{2}}\, \right. \nn \\ 
                 &\times&\left.   \Big[\phiDA_{\pi p}(\tau) 
                  - \frac{i}{6}\sigma_{\mu\nu}\,\frac{q'{}^\mu k'{}_j^\nu}{q'\cdot k'_j}
                            \,\phiDA'_{\pi\sigma}(\tau)
                + \frac16 \sigma_{\mu\nu}\,q'{}^\mu\,\phiDA_{\pi\sigma}(\tau) 
                            \frac{\partial}{\partial k_{\perp\nu}} \Big] \right\}_{k_\perp\to 0}
\label{eq:projector-2}
\ea
the subprocess amplitudes for the twist-2 and for the 2-particle 
twist-3 contributions have already been calculated in \ci{signatures}. 
The usual twist-2 pion \da{} is denoted by $\phiDA$ and $\phiDA_{p}$, 
$\phiDA_{\sigma}$ are the two 2-particle twist-3 \da s while  $\phiDA'_{\sigma}=d\phiDA_{\sigma}/d\tau$. 
Their definitions are given in App.\ A. 
In \req{eq:projector-2} $f_\pi$ is the familiar decay constant of the meson ($f_\pi=0.132\,\mev$); 
$\tau$ denotes the momentum fraction the quark entering the meson carries; $N_C$ the number of colors 
and $f$ and $g$ represent color labels of the quark and antiquark, respectively. The Dirac labels are 
omitted for convenience. In \req{eq:projector-2}, $k_{\perp}$ denotes the 
intrinsic transverse momentum of the quark entering the meson, defined with respect to the meson's 
momentum, $q'$. It is usually neglected in the collinear hard-scattering approach. The quark and
antiquark momenta are thus given by
\be
k_q\=\tau q' +k_\perp\,, \qquad k_{\bar{q}}\=\taub q'-k_\perp
\ee
where $\taub=1-\tau$. After the derivative in \req{eq:projector-2} is performed the collinear 
limit, $k_{\perp}=0$, is to be taken. Finally, the mass parameter $\mu_\pi$ is defined in Eq.\ 
\req{eq:mass-parameter}. Taking from \ci{PDG} the current-quark masses appearing in Eq.\ 
\req{eq:mass-parameter}, one obtains $\mu_\pi(\muO)=2.6\,\gev$ at the scale $\muO=4\gev^2$. 
The uncertainty of $\mu_\pi$ is however large~\footnote{
    For instance values of $1.8\,\gev$ and $1.9\,\gev$ at $\muO$ are quoted in \ci{huang} and 
    \ci{ball98}, respectively.}.   
The mass parameter evolves as
\be
\mu_\pi(\muR) \= L^{-4/\beta_0} \mu_\pi(\muO)
\ee
where 
\be
L\=\frac{\alpha_S(\muR)}{\alpha_S(\muO)}
  =\frac{\ln{(\muO/\Lambda^2_{\rm QCD})}}{\ln{(\muR/\Lambda^2_{\rm QCD})}} 
\ee
and $\beta_0=(11 N_C-2n_f)/3$. We work with four flavors ($n_f=4$) 
and adopt the value $\Lambda_{\rm QCD}=0.22\,\gev$. 
For the factorization and renormalization scale we choose 
$\muF=\muR$ and
\be
\muR\=\frac{\th \uh }{\sh}\,,
\ee
which takes care of the requirement that both t and u
should be large.

The twist-2 contribution only affects the subprocess amplitude for quark helicity non-flip.
At leading-order (LO) of perturbative QCD it reads \ci{signatures}
\be
{\cal H}^{twist-2}_{0\lambda,\mu\lambda}\=  
               2\pi\als(\muR) f_\pi \,\frac{C_F}{N_C}\,\frac{\sqrt{-\th/2}}{\uh\sh}\,   
             \langle 1/\tau \rangle_\pi \big[(1+2\lambda\mu)\sh - (1-2\lambda\mu)\uh\big]
\label{eq:twist2-sub}
\ee
where as usual $C_F=(N_C^2-1)/(2 N_C)$ is a color factor, while $\langle 1/\tau \rangle$ is the 
$1/\tau$ moment of twist-2 pion \da{}. The symmetry of twist-2 pion \da{} under the replacement
$\tau \leftrightarrow \taub$ is in \req{eq:twist2-sub} already taken into account.
For this \da{} we use the truncated Gegenbauer expansion
\be
\phiDA_\pi(\tau,\muR)\=6\tau\taub\,\big[1 + a_{2}(\muO)L^{\gamma_2/\beta_0}\; 
C_2^{3/2}(2\tau -1)\big]
\label{eq:pionDA}
\ee 
with the recent lattice QCD result on the second Gegenbauer coefficient \ci{braun15} 
\be
a_{2}(\muO)\=0.1364 \pm 0.0213
\ee
and the anomalous dimension $\gamma_2=50/9$. The $1/\tau$ moment of the twist-2 pion \da{} is 
given by
\be
\langle 1/\tau \rangle_\pi \= 3\big[1+a_{2}(\muR)\big]\,.
\ee

Let us turn to twist-3 contributions. The 2-particle twist-3 contributions were determined in
\ci{signatures}, and in this work we rewrite them in a compact  
form suitable for combining with 3-particle results. We list
both 2- and 3-particle twist-3 contribution in App. B, while, 
as we will show, their sum can be simplified and expressed 
in terms of the convolution with just the 3-particle twist-3 \da{}.
Typical lowest order Feynman diagrams relevant for 3-particle twist-3 contributions are shown 
in Fig.\ \ref{fig:graphs}. The 16 Feynman graphs with (c) and without (b) the triple-gluon 
coupling make up the 3-particle contribution. The graphs c and b have different color factors. 
Graphs of type d) for which the constituent gluon of the pion couples to one of its quark 
constituents, are  soft contributions and are to be considered as part of the meson wave function. 
In perturbation theory the gluon field, $A_\mu^a(x)$, appears in the vacuum-meson matrix elements
whereas the 3-particle \da{}, $\phiDA_{3\pi}$, is defined through the gluon field strength tensor, 
$G_{\mu\nu}$, see \req{eq:3-particle}. In light-cone gauge which we are using, the two quantities 
are related to each other by \ci{kogut-soper} 
\be 
A^a_\mu(z) \=\lim_{\eps\to 0} n^\nu \int_0^\infty d\sigma e^{-\eps\sigma} 
                                              G^a_{\mu\nu}(z+ n\sigma)
\label{eq:A-G}
\ee
where $n$ is a light-like vector with $n\cdot A=0$.
By making use of this relation and the definition of the 3-particle twist-3 \da{}
\req{eq:3-particle} we derive the expression for the vacuum-pion matrix element to be used
in the perturbative calculation of the contribution 
involving $q\bar{q}g$ Fock component
\ba
\langle 0| \bar{u}^{\,g}(z_b)A^{\beta,c}( z_g)d^f(z_a)|\pi^-(q')\rangle &=& 
  \int [d\tau]_3 \, e^{-i q' \cdot(\tau_a z_a+\tau_b z_b+\tau_g z_g)} 
{\cal P}_{3,fg}^{\beta,c} \,.
\label{eq:da-projector3}
\ea  
with the 3-particle projector, $q\bar{q}g\to \pi$,
given by
\be
{\cal P}_{3,fg}^{\beta,c}\=
             \frac{i}{g}\,\frac{f_{3\pi}}{2 \sqrt{2 N_C}}\, \frac{\left(t^c\right)_{fg}}{C_F\sqrt{N_C}}\, 
           \frac{\gamma_5}{\sqrt{2}} \, \sigma_{\mu\nu} q'{}^{\mu} g_\perp^{\nu\beta}\, 
  \frac{\phiDA_{3\pi}(\tau_a,\tau_b,\tau_g)}{\tau_g} 
\,. 
\label{eq:projector-3}
\ee
The transverse metric tensor is defined as~\footnote{
    We remind the reader that in our symmetrical c.m.s. the pion and the outgoing quark move
    back to back, i.e.\ $\vec{k}'_j=-\vec{q}'$. Transforming $q'$ and $k'_j$ to a frame in which 
    the pion moves along the 3-direction these momenta become
\be
    q' \to [q^+,0,\vec{0}_\perp]\,, \qquad k'_j\to [0, q^+,\vec{0}_\perp]\,,  \nn
\ee
   with the pion mass being neglected. In this frame the tensor $g^{\nu\beta}_\perp$
   has the components $g_\perp^{11}=g_\perp^{22}$ while all other components are zero.}
\be
g^{\nu\beta}_\perp\= \left(g^{\nu\beta}-\frac{k_j'{}^\nu q'{}^\beta 
                                     + q'{}^\nu k_j'{}^\beta}{k'_j\cdot q'}
                          \right)\,,
\label{eq:transverse-metric-tensor}
\ee
and the integration measure, $[d\tau]_3$, is defined in \req{eq:measure}, while $t^c=\lambda^c/2$ 
is the SU(3) color matrix for a gluon of color $c$ and $g$ denotes the QCD coupling. 
As is detailed in App.\ A the equation of motion relates the 2- and 3-particle twist-3 \da s to each
other. In light-cone gauge the relation for the antiquark for instance reads (see \req{eq:EOM2} and 
\req{eq:phi-eom2})
\be
f_\pi\mu_\pi\Big[\taub\phiDA_{\pi p}(\tau) - \frac16 \taub \phiDA_{\pi\sigma}'(\tau) 
             - \frac13 \phiDA_{\pi\sigma}(\tau) \Big]
                    \= 2 f_{3\pi}\,\int_0^{1-\tau} \frac{d\tau_g}{\tau_g} 
                               \phiDA_{3\pi} (\tau, \taub-\tau_g, \tau_g)\,.
\label{eq:relation1}
\ee 
Using this relation, we can express the full twist-3 subprocess amplitude, the sum of the 2-particle
and 3-particle contributions, through the 3-particle \da{} alone   
\ba
{\cal H}^{twist-3}_{0-\lambda,\mu\lambda}&=&  
               4\pi\als(\muR)\,\frac{f_{3\pi}(\muR)}{N_C}\,(2\lambda-\mu)
              \sqrt{-\frac{\uh\sh}{2}}\,\frac1{\sh^2\uh^2} \nn\\
  &\times& \int_0^1 d\tau \int_0^{\taub} \frac{d\tau_g}{\tau_g} 
                             \phiDA_{3\pi}(\tau,\taub-\tau_g,\tau_g,\muR) 
\nn\\
      &\times& \left[C_F\Big(\frac1{\taub^2} - \frac1{\taub(\taub-\tau_g)}\Big)\Big(\sh^2+\uh^2\Big)
                 \right. \nn\\
   &+& \left.\Big(C_F-\frac12 C_A\Big)\Big(\frac1{\tau} + \frac1{\taub-\tau_g}\Big)\,
                 \frac{\th^2}{\tau_g} \,\right]
\label{eq:twist3-sub}
\ea 
where $C_A=N_C$. The first term in \req{eq:twist3-sub} represents a combination of 2 and 3-particle 
contributions while the second term is a pure 3-particle contribution. For the interested reader 
we present the 2- and 3-particle contributions separately in App.\ B. We see 
from \req{eq:twist3-sub} that the twist-3 contribution only feeds the quark helicity-flip subprocess 
amplitudes in contrast to the twist-2 contribution which controls the helicity non-flip ones. 

We have checked our results by analyzing the gauge invariance conditions.
As expected the 2- and 3-particle twist-3 contributions are separately gauge invariant with respect 
to the choice of gauge of the virtual gluon. In contrast, the gauge invariance with respect to the 
choice of gauge of the photon is only satisfied by the complete twist-3 result but not separately 
for the 2- and 3-particle contributions. Another important property of the amplitude 
\req{eq:twist3-sub} is its crossing symmetry. As has been shown long ago \ci{CGLN} the amplitudes 
${\cal H}_{0-\lambda,\mu\lambda}$ are $\sh - \uh$ crossing symmetric to any order of perturbation 
theory~\footnote{ 
     The crossing behavior of the amplitude ${\cal H}_{0\lambda,\mu\lambda}$ is more complicated 
     but, as shwon in \ci{signatures}, the expression \req{eq:twist2-sub} has the correct 
     $\sh - \uh$ crossing property.}
which is evidently the case for \req{eq:twist3-sub}. As can be seen in App.\ B the separate 
2- and 3-particle contributions are not crossing symmetric. Thus, as is evident from the above
remarks, both, the 2- and 3-particle twist-3 contributions have to be taken into account 
in order to obtain a physically consistent result that respects the fundamental properties of
gauge invariance and crossing symmetry.

Obviously the 2-particle twist-3 contribution
vanishes if the 3-particle \da{} is assumed to be zero.
This has already been noticed in \ci{signatures}.
This situation is to be contrasted with that one in deeply virtual electroproduction of 
pseudoscalar mesons. In the latter process the contribution from the solution of 
\req{eq:2particle-det} for $\phiDA_i^{\rm EOM}=0$ - the so-called Wandzura-Wilczek approximation -
\be
\phiDA^{WW}_p\equiv 1\,, \qquad \phiDA^{WW}_\sigma\=6\tau\taub
\label{eq:WW}
\ee 
does not vanish. Up to corrections of order $t/Q^2$ where $Q^2$ is the virtuality of the photon,
the subprocess amplitude in electroproduction is under control of the \da{} $\phiDA_{\pi p}$. Because
of its end-point behavior it leads to an infrared singularity in collinear approximation.
In \ci{GK5,GK6} this singularity is regularized by retaining the quark transverse
momenta in the subprocess.

The 3-particle \da{} can be expanded upon the Jacobi polynomials \ci{braun90}. We employ a truncated
version of it:
\ba
\phiDA_{3\pi}(\tau_a,\tau_b,\tau_g,\muR)&=& 360\tau_a\tau_b\tau_g^2\Big[1 
                  + \omega_{1,0}(\muR)\frac12(7\tau_g-3)\nn\\
             &+& \omega_{2,0}(\muR)(2-4\tau_a\tau_b-8\tau_g+8\tau_g^2)   \nn\\
             &+& \omega_{1,1}(\muR)(3\tau_a\tau_b-2\tau_g+3\tau_g^2)\Big]\,.
\label{eq:3-particle-da}
\ea
The parameters of the 3-particle \da{} evolve as:
\ba
f_{3\pi}(\muR)&=& L^{(16/3C_F-1)/\beta_0}\;f_{3\pi}(\muO)\,, \nn\\ 
\omega_{1,0}(\mu_R)&=& L^{(-25/6C_F+11/3C_A)/\beta_0}\omega_{1,0}(\muO)\,,  \nn\\
\omega_{11}(\muR) &=& \frac1{\gamma_+-\gamma_-}\,
            \left[(\gamma_--\gamma_{qq}) A_+(\muO) L^{(\gamma_+-16/3 C_F+1)/\beta_0} \right. \nn\\
            &&\left. \hspace*{0.12\tw}  
            + (\gamma_+-\gamma_{qq}) A_-(\muO) L^{(\gamma_--16/3 C_F+1)/\beta_0}\right]\,,  \nn\\ 
\omega_{20}(\muR) &=& \frac14\frac{\gamma_{qg}}{\gamma_--\gamma_+}\,
    \left[A_+(\muO) L^{(\gamma_+-16/3C_F+1)/\beta_0}  \right. \nn\\
           && \left. \hspace*{0.13\tw} + A_-(\muO) L^{(\gamma_--16/3C_F+1)/\beta_0}\right]\,,
\ea
where  
\ba
A_+(\muO)&=& -\omega_{11}(\muO) - 4\frac{\gamma_+-\gamma_{qq}}{\gamma_{qg}}\omega_{20}(\muO)\,, \nn\\
A_-(\muO)&=& \phantom{-}\omega_{11}(\muO) + 4\frac{\gamma_--\gamma_{qq}}
                                             {\gamma_{qg}}\omega_{20}(\muO)\,.   
\ea
The anomalous dimensions are
\be
\gamma_{qq}\=\frac{122}{9}\,, \quad \gamma_{gg}\=\frac{511}{45}\,, \quad \gamma_{qg}\=\frac53\,,
\quad \gamma_{gq}\=\frac{21}{5}\,,
\ee
with the eigenvalues
\be
\gamma_\pm \= \frac12\big[\gamma_{qq}+\gamma_{gg}
               \pm\sqrt{(\gamma_{qq}-\gamma_{gg})^2+4\gamma_{qg}\gamma_{gq}}\;\Big]\,.
\ee
The anomalous dimensions are to be found in the literature \ci{braun90,ball98}.

With the help of \req{eq:3-particle-da} the integrations in \req{eq:twist3-sub} can be performed
analytically
\ba
{\cal H}^{twist-3}_{0-\lambda,\mu\lambda}&=&  
               -80\pi\als\frac{f_{3\pi}}{N_C}\,(2\lambda-\mu)
              \sqrt{-\frac{\uh\sh}{2}}\,\frac1{\sh^2\uh^2} \nn\\
            &\times& \Big[ C_F \big(1-\frac3{16}\omega_{1,0} 
                     + \frac6{25}\omega_{2,0} - \frac3{50}\omega_{1,1}\big)
                       \big(\sh^2+\uh^2\big) \nn\\
      &-& \big(C_F -\frac12 C_A\big)\big(6-\frac{15}{4}\omega_{1,0} + \frac{12}{5}\omega_{2,0}
                     + \frac35\omega_{1,1}\big)\,\th^2\, \Big]\,.
\ea

\section{Predictions for photoproduction of pseudoscalar mesons}
\label{sec:pi0-results}
\subsection{The cross section for $\pi^0$ photoproduction}
The most recent determination of the 3-particle pion \da{} is made in \ci{ball98} on the basis of
QCD sum rules. Instead of $f_{3\pi}$ the parameter $\eta_3$ is quoted in that work. It is related 
to $f_{3\pi}$ by \req{eq:eta3-def}. Evolved to the scale $\muO$ the value of $\eta_3$ derived in 
\ci{ball98} leads to
\be
f_{3\pi}(\muO)\=0.004\,\gev^2\,.
\label{eq:f3pi}
\ee
The expansion coefficients of the 3-particle \da{} quoted in \ci{ball98} are
\be
\omega_{1,0}(\muO)\=-2.55\,, \quad \omega_{2,0}(\muO)\=\omega_{1,1}(\muO)\=0\,.
\label{eq:omega}
\ee
According to \ci{ball98}  the uncertainties of the parameters \req{eq:f3pi} and \req{eq:omega} 
are large, of order of $30\%$. 

We are now in the position to evaluate the photoproduction cross section defined by
\be
\frac{d\sigma}{dt} \= \frac1{32\pi (s-m^2)^2}\Big[ |{\cal M}_{0+++}|^2
   + |{\cal M}_{0+-+}|^2 + |{\cal M}_{0-++}|^2 + |{\cal M}_{0--+}|^2\Big]\,.
\label{eq:cross-section}
\ee
\begin{figure}[t]
\begin{center}
\includegraphics[width=0.59\tw]{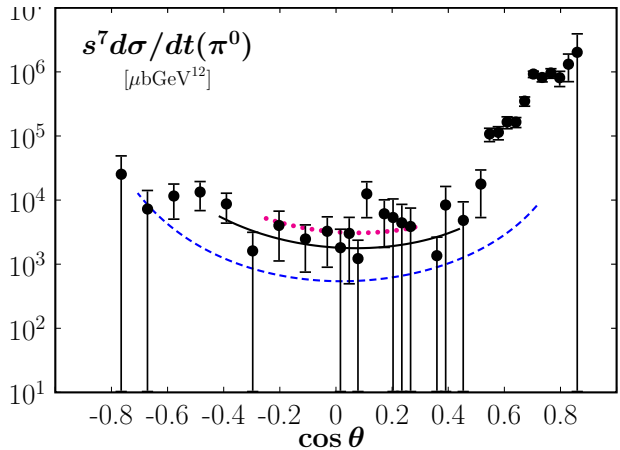}
\caption{\label{fig:dsdt} Results for the cross section of $\pi^0$ photoproduction versus
the cosine of the c.m.s. scattering angle, $\theta$. The solid (dashed, dotted) curves represent 
our results at $s=11.06 \,(20, 9)\gev^2$. The data at $s=11.06\,\gev^2$ are taken from CLAS 
\ci{clas-pi0}. The cross sections are multiplied by $s^7$  and the theoretical results are only 
shown for $-t$ and $-u$ larger than $2.5\,\gev^2$.}
\end{center}
\end{figure} 
It turns out that with the 3-particle \da{} specified in Eqs.\ \req{eq:f3pi} and \req{eq:omega}
the cross section for $\pi^0$ photoproduction is still somewhat small as compared to the
CLAS data \ci{clas-pi0}. 
Since there is no physical reason why $\omega_{20}$ should be zero and its contribution is by no 
means suppressed as compared to $\omega_{10}$ we fit this parameter to the CLAS data. We obtain
\be
\omega_{20}(\muO)\=8.0\,.
\label{eq:omega20}
\ee
This value is a bit smaller than the value quoted in \ci{braun90,chernyak}. The results of the 
fit to the $\pi^0$ cross section are shown in Fig.\ \ref{fig:dsdt}. The cross section is 
multiplied by $s^7$. This scaling behavior which holds at a fixed c.m.s. scattering angle 
$\theta$, follows from dimensional counting for the leading-twist contribution. In order to 
match roughly the requirement for the handbag approach of Mandelstam variables much larger 
than $\Lambda^2$ we only show results for $-t$ and $-u$ larger than $2.5\,\gev^2$. As one sees 
from Fig.\ \ref{fig:dsdt} our results are in reasonable agreement with the CLAS data \ci{clas-pi0} 
at $s=11.06\,\gev^2$. For comparison we also present predictions at $s=9$ and $20\,\gev^2$. 
Obviously, the theoretical results drop faster with energy than $s^{-7}$. Leaving aside the logs 
of $s$ from the evolution the leading-twist handbag results would scale as $s^{-7}$ only if the 
form factors $R_V$ and $R_A$ would drop as $1/t^2$ which is not exactly the case, see Tab.\ 
\ref{tab:2}. In the range of $s$ we are interested in, our cross section effectively behaves 
$\propto s^{-9}$. This is a consequence of the twist-3 dominance. From the subprocess amplitude 
one gets a suppression factor $\mu_\pi^2/s$  in the cross section as compared to the twist-2 
contribution, cf.\ \req{eq:twist2-sub} and \req{eq:twist3-sub}. In addition there are the logs 
of $s$ from the evolution of the \da s. The transversity form factors effectively contribute to the 
energy dependence of the cross section somewhat stronger than  $1/s^4$ because their stronger 
$t$-dependence (see Tab.\ \ref{tab:2}) is only partly compensated by the extra factors of $t$ in 
the amplitudes \req{eq:amplitudes}. Since our form factors represent $1/x$-moments of GPDs they 
evolve with the scale in principle. This effect is neglected by us for the following reason:
because of the strong $x-t$ correlation the form factors at large $-t$ are under control of a narrow
region of large $x$. With increasing $-t$ this region approaches 1. Therefore, our form factors 
approximately become equal to the scale-independent lowest moment of the GPDs concerned (e.g.\ 
$R_V^a\to F_1^a$ for $-t\to \infty$ where $F_1^a$ is the flavor-$a$ Dirac form factor of the proton).
Thus, as it is argued in \ci{DFJK4}, the $1/x$-factors in the form factors can be viewed as a 
phenomenological estimate of effects beyond the strict $\Lambda/\sqrt{-t}$ expansion. 
If the scale-dependence of the form factors is neglected one may also ignore that of the 
\da s~\footnote{
  Taking the parameters $f_{3\pi}=0.005\,\gev^2$, $\omega_{1,0}=-3$,
  $\omega_{2,0}=7$ and $\omega_{1,1}=0$, valid at a low scale
  of about $1\,\gev$ and ignoring evolution, the predictions for the photoproduction
  cross section at $s=11.06\,\gev^2$ is almost indistinguishable
  from that one shown in Fig. 3.} 
. Since at fixed $s$ the renormalization scale (24) scarcely varies in the wide-angle
region the shape of the theoretical results on the cross section is hardly altered in this case
but the effective energy dependence of $d\sigma/dt$ is reduced to about $s^{-8}$.
In contrast to electroproduction of the $\pi^0$ \ci{GK6} the cross section is not dominated by
a single transversity GPDs but the total twist-3 contribution do. For $\pi^0$ 
electroproduction the twist-3 contribution feeds the cross section for transversely
polarized photons while twist-2 controls the longitudinal one. Hence, twist-3 dominance
means the dominance of the transverse cross section in $\pi^0$ electroproduction which is 
experimentally confirmed \ci{defurne} for photon virtualities of order of $2\,\gev^2$. In 
$\pi^0$ photoproduction, on the other hand, both twist-2 and twist-3 contribute to the same 
helicity amplitudes leading to interference terms in the cross section. The twist-2-twist-3 
interference term is largest in the forward hemisphere. For $-t\to 2.5\,\gev^2$ it is negative 
and amounts to $10 - 15\%$ in absolute value. In the backward hemisphere the interference term 
amounts to merely a few per cent.  Since the twist-3 contribution dominates, the uncertainty of 
our cross section is correspondingly large. In fact, the parametric uncertainty of the cross 
section arising from those of the transversity form factors and the twist-3 \da{}, is about 
$70\%$ near 90 degrees. 
   
As we discuss in App.\ A, the 3-particle twist-3 \da{} fixes the  2-particle twist-3  \da s 
through the equations of motion. For the \da{} \req{eq:3-particle-da} with the parameters
\req{eq:f3pi}, \req{eq:omega} and \req{eq:omega20} the Gegenbauer coefficients of the 
 2-particle twist-3  \da s  are (see Eqs.\ \req{eq:P-gegenbauer}, \req{eq:S-gegenbauer} and 
\req{eq:S-norm})
\ba
a^p_{\pi 2}(\muO)&=&- 0.56\,,\phantom{0} \qquad a^p_{\pi 4}(\muO)\=0.17\,, \nn\\
a^{\sigma}_{\pi 2}(\muO)&=&-0.084\,, \qquad a^{\sigma}_{\pi 4}(\muO)\=0.031\,, 
\label{eq:a-numerical}
\ea
($a^p_{\pi n}=a^{\sigma}_{\pi n}=0$ for $n\geq 4$) and
\be
\eta_\sigma(\muO)\=0.64\,.
\ee 
The values of the Gegenbauer coefficients $a^p_{\pi 2}$ and, with regard to the value of 
$\eta_\sigma$, also that of $a^\sigma_{\pi 2}$ are compatible with those to be found in the 
literature while the coefficients $a^{p(\sigma)}_{\pi 4}$ have opposite sign. These 2-particle 
twist-3 \da s from the literature have been derived with various methods: the Dyson-Schwinger 
approach \ci{shi15}, a light-cone quark model \ci{choi17} and a chiral quark model \ci{nam06}. 
The 3-particle \da{} is not considered in these papers and therefore no result on $f_{3\pi}$ is 
quoted. However, this parameter plays an important role in the present work.  
Our values for the Gegenbauer coefficients of $\phiDA_{\pi p}$ \req{eq:a-numerical} have opposite 
signs to those quoted in \ci{ball98}. The latter Gegenbauer coefficients have been derived from 
the same 3-particle \da{} that we are using but the Fock-Schwinger gauge is employed in the 
vacuum-meson matrix elements. Thus, the different methods applied in \ci{ball98} and by us lead 
to drastic differences in the 2-particle twist-3 \da s.  The normalization $\eta_\sigma$ may be 
absorbed into, say, the mass parameter $\mu_\pi$ in the case of the \da{} $\phiDA_{\pi\sigma}$ 
leading to a mass parameter $\mu_{\pi\sigma}$ which is somewhat smaller than the mass parameter 
$\mu_\pi$ appearing for $\phiDA_{\pi p}$. In \ci{huang} it is claimed that such differences in the 
mass parameter may be generated by the off-shellness of the quarks and antiquarks in the pion.

\subsection{Spin effects}
The derivation of the photoproduction amplitudes within the handbag approach naturally
requires the use of the light-cone helicity basis. However, for comparison with experimental
results on spin-dependent observables, the use of ordinary photon-proton c.m.s. helicity basis
is more convenient. The standard helicity amplitudes, $\Phi_{0\nu',\mu\nu}$
are obtained from the light-cone helicity amplitudes \req{eq:amplitudes}, 
by the transform \ci{diehl01}
\be
\Phi_{0\nu',\mu\nu}\= {\cal M}_{0\nu',\mu\nu}
                   +\frac12\kappa\Big[ (-1)^{1/2-\nu'}\,{\cal M}_{0-\nu',\mu\nu}
                   +(-1)^{1/2+\nu}\,{\cal M}_{0\nu',\mu-\nu}\Big] + {\cal O}(m^2/s)
\ee
where
\be
\kappa\=\frac{2m}{\sqrt{s}}\,\frac{\sqrt{-t}}{\sqrt{s}+\sqrt{-u}}\,.
\ee
For convenience the notation for the helicities is kept.  
Obviously,
\be
\sum_{\nu',\mu}|\Phi_{0\nu',\mu +}|^2\=\sum_{\nu',\mu}|{\cal M}_{0\nu',\mu +}|^2\,. 
\ee

As for wide-angle Compton scattering \ci{kroll17,HKM} the most interesting spin-dependent 
observables are the correlations of the helicities of the incoming photon and the incoming, 
$A_{LL}$, or 
outgoing proton, $K_{LL}$:
\ba
A_{LL} &=& \frac{|\Phi_{0+,++}|^2 - |\Phi_{0+,-+}|^2 + |\Phi_{0-,++}|^2 - |\Phi_{0-,-+}|^2 }
                {\sum_{\nu',\mu}|\Phi_{0\nu',\mu +}|^2}\,, \nn\\
K_{LL} &=& \frac{|\Phi_{0+,++}|^2 - |\Phi_{0+,-+}|^2 - |\Phi_{0-,++}|^2 + |\Phi_{0-,-+}|^2 }
                {\sum_{\nu',\mu}|\Phi_{0\nu',\mu +}|^2}\,. 
\ea
One can easily check that for the twist-3 contribution one has 
\be
A_{LL}^{twist-3}\=-K_{LL}^{twist-3}
\ee
while for the twist-2 contribution
\be
A_{LL}^{twist-2}\=\phantom{-}K_{LL}^{twist-2}
\ee
holds as is the case for wide-angle Compton scattering \ci{kroll17,HKM}. Thus, the helicity 
correlations may provide a characteristic signal for the dominance of twist-3 contribution in 
photoproduction of pseudoscalar mesons. Thus the observables $A_{LL}$ and $K_{LL}$ play a similar 
important role for the discrimination between twist-2 and twist-3 in photoproduction of pions as 
the longitudinal and transverse cross sections in pion electroproduction.

In terms of helicity amplitudes the correlation between the helicity of the photon and the
sideway polarization of the incoming proton is~\footnote{
  Sideway is defined as the direction perpendicular to the proton momentum but in the scattering 
  plane.}
\be
A_{LS}\= 2 \frac{{\rm Re}\Big[\Phi^{*}_{0+,++}\Phi_{0-,-+} - \Phi^{*}_{0+,-+}\Phi_{0-,++}\Big]}
                 {\sum_{\nu',\mu}|\Phi_{0\nu',\mu +}|^2} 
\ee
and the correlation between the helicity of the photon and the sideway polarization of the 
recoil proton 
\be
K_{LS}\= 2 \frac{{\rm Re}\Big[\Phi^{*}_{0+,++}\Phi_{0-,++} - \Phi^{*}_{0+,-+}\Phi_{0-,-+}\Big]}
                 {\sum_{\nu',\mu}|\Phi_{0\nu',\mu +}|^2}\,. 
\ee
The last spin observable we consider is the asymmetry for linearly polarized photons, transverse and
parallel to the photon momentum,
\be
\Sigma\=2\frac{{\rm Re}\Big[\Phi^{*}_{0+,++}\Phi_{0+,-+} + \Phi^{*}_{0-,++}\Phi_{0-,-+}\Big]}
                           {\sum_{\nu',\mu}|\Phi_{0\nu',\mu +}|^2}\,. 
\ee
Since the twist-3 subprocess amplitude, ${\cal H}^{twist-3}_{0-,++}$, is zero as can be seen from 
\req{eq:twist3-sub}, any spin observable is only given by a ratio of the transversity form factors 
up to corrections from twist-2. Hence, the predictions on spin-dependent observables are more 
precise than those for the cross sections since only the uncertainties of the form factors matter. 
Consequently, they do not suffer from the large uncertainties arising from the 3-particle
\da{} as is the case for the differential cross section. 

It is instructive to quote the observables obtained from the twist-3 contribution alone since 
this is the dominant contribution. In this case the cross section is given by
\be
\frac{d\sigma^{twist-3}}{dt}\=\frac{\pi\ale}{32(s-m^2)^2} |{\cal H}^{twist-3}_{0-,-+}|^2 F^{\pi^0}
\ee
where the combination of form factors, $F^{\pi^0}$, reads
\be
F^{\pi^0}\=-\frac{t}{2m^2}\Big[ \bar{S}_T^{\pi^0 2} 
                          - \frac{t}{m^2}S_S^{\pi^0 2} + 4 S_S^{\pi^0} S_T^{\pi^0} 
                                      -8\frac{m^2}{t}S_T^{\pi^0 2}\Big]\,.
\ee
The spin-dependent observables then read
\ba
A_{LL}^{twist-3} &=& -K_{LL}^{twist-3} \=- 4 \frac{S_T^{\pi^0}
                          \Big[S_T^{\pi^0}-\frac{t}{2m^2}S_S^{\pi^0} 
                    +\kappa \frac{\sqrt{-t}}{2m}\bar{S}_T^{\pi^0}\Big]}{F^{\pi^0}}\,, \nn\\
A_{LS}^{twist-3} &=& -K_{LS}^{twist-3} \= -2 \frac{S_T^{\pi^0}}{F^{\pi^0}}
     \Big[ \frac{\sqrt{-t}}{m}\;\bar{S}_T^{\pi^0} -2\kappa(S_T^{\pi^0}
                        -\frac{t}{2m^2}S_S^{\pi^0})\Big]\,, \nn\\
\Sigma^{twist-3} &=& 1 - 4 \frac{S_T^{\pi^0 2}}{F^{\pi^0}}\,.                           
\label{eq:spin}
\ea
Since only the form factors are needed in \req{eq:spin}, it seems possible
to give predictions of such observables for different meson channels. It is also evident from 
\req{eq:spin} that these spin observables are independent on energy at fixed $t$ up to corrections 
from twist-2 and corrections of order $\Lambda^2/s$.

\begin{figure}[t]
\begin{center}
\includegraphics[width=0.59\tw]{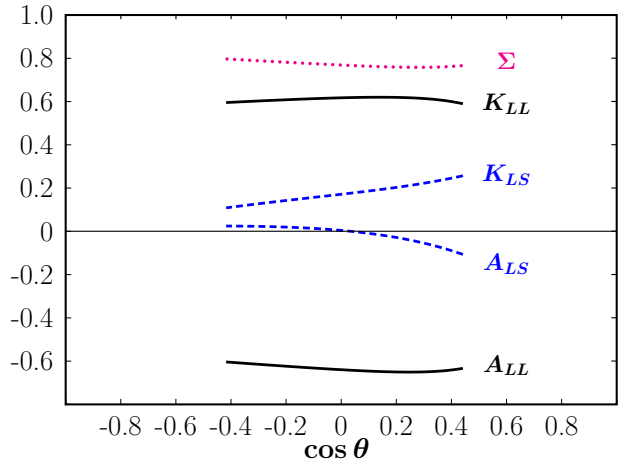}
\caption{\label{fig:spin} Predictions for spin observables of $\pi^0$ photoproduction
at $s=11.06\,\gev^2$. The parametric uncertainty is $\simeq 15 \%$ near 90 degrees.}
\end{center}
\end{figure}

In Fig.\ \ref{fig:spin} we show predictions on the spin-dependent observables for $\pi^0$ 
photoproduction. One sees that $A_{LL}$ and $K_{LL}$ are large in absolute value and almost mirror 
symmetrical. The observables $A_{LS}$ and $K_{LS}$ are small in absolute value. The twist-2 
contributions to them are relatively large. The observable $\Sigma$ is close to unity and only 
mildly $t$-dependent. In tendency this is in agreement with a glueX measurement for $\pi^0$ 
photoproduction at small $-t$ \ci{glueX-sigma}.

At Jefferson Lab the observables $K_{LL}$ and $K_{LS}$ have been measured twice: at
$s=7.8\,\gev^2$ and a c.m.s. scattering angle of $70$ degrees \ci{fanelli} and at $s=6.9\,\gev^2$
and $\theta=120^\circ$ \ci{hamilton}. These kinematical settings do not respect the 
requirement of large Mandelstam variables, either $-t$ or $-u$ is too small. The data are as
follows:
\ba
\ci{fanelli}\quad\; s&=&7.8\,\gev^2\,, \qquad t=-2.1\phantom{0}\,\gev^2\,:   \hfill \nn\\
               &&  K_{LL}\=-0.082\pm 0.007\,, \qquad K_{LS}\=-0.296\pm 0.007\,, \nn\\
\;\ci{hamilton}\quad\; s&=& 6.9\,\gev^2\,, \qquad u=-1.04\,\gev^2\,:  \hfill \nn\\
               && K_{LL}\=\phantom{-}0.532\pm 0.006\,, \qquad K_{LS}\=\phantom{-}0.480\pm 0.007\,.
\ea
Inspection of Fig.\ \ref{fig:spin} shows that the data at $6.9\,\gev^2$ agree
with our predictions in tendency while the $7.8\,\gev^2$ data are smaller. This is similar to 
the situation in wide-angle Compton scattering \ci{kroll17}. Before conclusions can be drawn 
we have to wait for data at $s, -t, -u\gg \Lambda^2$. Such data are planned to measure at 
Jefferson Lab. \ci{bogdan}.
\subsection{Other channels} 

\begin{figure}[t]
\begin{center}
\includegraphics[width=0.59\tw]{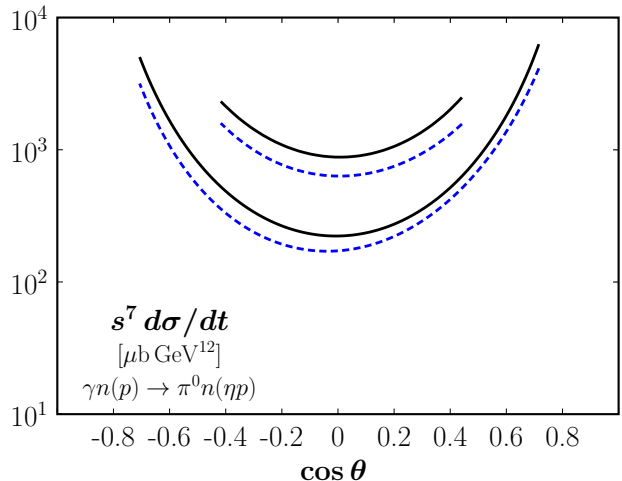}
\caption{\label{fig:eta} Predictions for the cross sections of $\pi^0$ photoproduction off 
neutrons (solid lines) and $\eta$ photoproduction (dashed lines) at at $s=11.06\,$ (upper lines) 
and $20\gev^2$ (lower lines). The parametric uncertainty amounts to about $70\%$ near 90 degrees.}
\end{center}
\end{figure}

In this section we are going to comment briefly on other wide-angle photoproduction processes.
From the theoretical point of view the simplest case is of course $\pi^0$ photoproduction 
off neutrons. In this case we only have  to change the process form factors. By isospin 
invariance the form factors are now
\be
F_{in}^{\pi^0}(t) \= \frac1{\sqrt{2}} \big[e_u F_i^d(t) - e_d F_i^u(t) \big]
\label{eq:pi0-neutron-FF}
\ee  
instead of \req{eq:pi0-form-factors}. Predictions for the corresponding cross section are
shown in Fig.\ \ref{fig:eta}. They are about a factor of 2.5 smaller than those 
for $\pi^0$ photoproduction off protons. For very large $-t$ the ratio of cross sections
for $\pi^0$ photoproduction off neutrons and off protons becomes equal to $(e_d/e_u)^2$ since, 
according to Tab.\ \ref{tab:2}, the $d$-quark form factors drop faster with increasing $-t$ than 
the $u$-quark ones. Consequently, the $d$-quark form factors can be neglected at large $-t$.
Spin effects are similar to those for the case of a proton target, see Fig.\ \ref{fig:spin}. 
The observables $A_{LL}, K_{LL}, A_{LS}$ and $K_{LS}$ are merely somewhat smaller in absolute value 
than the corresponding observables measured with a proton target.

\begin{figure}[t]
\begin{center}
\includegraphics[width=0.59\tw]{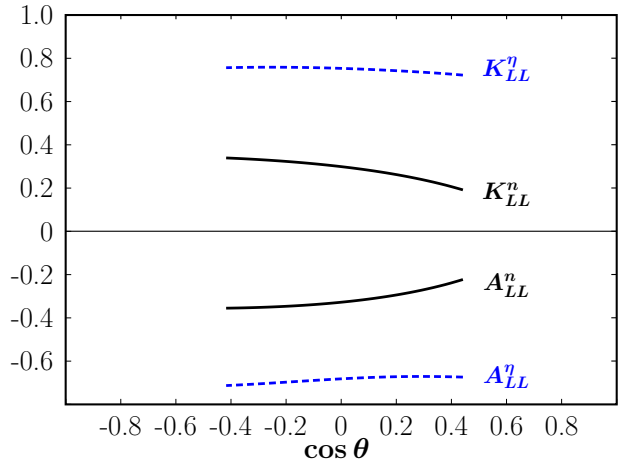}
\caption{\label{fig:spin-n-eta} Predictions for spin observables of $\pi^0$ photoproduction
off neutrons and of $\eta$ production at $s=11.06\,\gev^2$. The parametric uncertainty is 
$\simeq 15 \%$ near 90 degrees.}
\end{center}
\end{figure}

For $\eta$ photoproduction off protons the situation is more complex as is detailed in \ci{GK6}.
The mixing of the $\eta$ and $\eta'$ is to be taken into account and the form factors for 
strange quarks are also needed in principle. However, for the charge-conjugation even mesons 
the GPDs only contribute in the valence quark (or flavor non-singlet) combination 
$F_i^a-F_i^{\bar{a}}$. 
For the strange quark it seems to be plausible to assume $F^s_i\simeq F_i^{\bar{s}}$ \ci{DK13}. 
Hence, there is no contribution from strange quarks and, as is discussed in \ci{GK6}, 
the flavor-octet and singlet form factors are approximately given by
\be
F_{i}^{(8)}(t) \simeq \frac1{\sqrt{2}} F_i^{(1)}(t) \simeq 
                                \frac1{\sqrt{6}} \big[e_u F_i^u(t) + e_d F_i^d(t) \big]\,.
\label{eq:eta-FF}
\ee  
Using the mixing scheme advocated for in \ci{FKS1}, one can also decompose the $\eta$
amplitudes in a flavor-octet and single part
\be
{\cal M}_i^\eta\= \cos{\theta_8}{\cal M}_i^{(8)} - \sin{\theta_1} {\cal M}_i^{(1)}
\ee
with the mixing angles
\be
\theta_8\=-21.2^\circ\,, \qquad \theta_1\=-9.2^\circ\,,
\ee
derived in \ci{FKS1} on exploiting the divergences of the axial-vector current. Assuming
furthermore that the octet and singlet \da s, for both  twist-2 and twist-3, are the same
as the pion \da s and taking the values~\footnote{
  For the mass parameter $\mu_\eta^{(8)}$ one may approximately take the same value as for
  $\mu_\pi$ \ci{GK6}.}
\be
f_8\simeq 1.26 f_\pi\,, \ci{FKS1}   \qquad f_{3\eta}^{(8)}\=0.87 f_{3\pi}\,, \ci{ball98}
\ee
for the decay constants, one finds for the $\eta$ amplitude
\be
{\cal M}_i^\eta \simeq {\cal M}_i^{(8)} \big( \cos{\theta_8} - \sqrt{2}\sin{\theta_1}\big)\,.
\ee 
In principle there is also a contribution from the two-gluon Fock component of the $\eta$
\ci{two-gluon}. Since this contribution possesses flavor-singlet quantum numbers and is
of leading-twist nature, it can safely be neglected. Predictions for $\eta$ photoproduction 
cross section are shown in Fig.\ \ref{fig:eta}. The $\eta$ cross section is similar in shape  
to the $\pi^0$ one but about an factor of 2 smaller than the cross section for $\pi^0$ production 
off neutrons.

In Fig.\ \ref{fig:spin-n-eta} we present predictions on the helicity correlations, $A_{LL}$
and $K_{LL}$, for $\pi^0$ photoproduction off neutrons and for $\eta$ photoproduction off proton.

It is also possible to calculate observables for the photoproduction of charged pions and  
kaons. In these reactions transition GPDs and form factors appear which, due to
flavor symmetry, are related to the proton-proton ones \ci{frankfurt}. For instance the form factors
for photoproduction of charged pions are given by
\be
F_{i p\to n}^\pi\=F_{i n\to p}^\pi\=F_i^u - F_i^d\,,
\ee
Since this combination of flavor form factors is not well known we refrain from presenting 
predictions for the corresponding cross sections.
A new feature for the kaon-hyperon channels is the appearance of form factors for strange quarks 
which, at large $-t$, are unknown as yet. Moreover, the 3-particle \da{} for the kaon is also 
needed which is not well known \ci{ball06}. We therefore refrain from  giving predictions for 
these channels too. We stress, however, that from data on these channels one may extract 
information on the form factors and the twist-3 \da s. For instance, from spin-dependent 
observables one may learn about the form factors for strange quarks and subsequently from the 
differential cross section on the 3-particle \da{} of the kaon.

\section{Summary}
\label{sec:summary}

We have calculated wide-angle photoproduction of  $\pi^0$ mesons
within the handbag factorization scheme to twist-3 accuracy. 
The twist-3 contribution includes both the 2-particle,
$q\bar{q}$, as well as the 3-particle, $q\bar{q}g$, parts. 
In light-cone gauge which we are using for the vacuum-meson matrix elements, 
the equation of motion enables us to express the 2-particle twist-3 \da s through an 
integral upon the 3-particle \da. This relation is formally an inhomogeneous linear 
differential equation of first order which can readily be solved for a given 3-particle \da. 
The use of light-cone gauge made it also possible to derive a compact 3-particle twist-3 
projector in momentum space. Our twist-3 subprocess amplitude respects gauge invariance in
QCD and QED and $\sh - \uh$ crossing symmetry. On the other hand, the separate 2- and 
3-particle twist-3 amplitudes do not possess these properties. This reveals the necessity 
of taking into account both the 2- and 3-particle contributions in order to obtain physically 
consistent result. Our calculation method for the subprocess amplitudes is similar 
to a one exploited in \ci{wallon,anikin02} in a calculation of electroproduction of 
transversely polarized $\rho$-mesons.
We emphasize that the twist-3 effect we considered which follows from
the twist-3 pion DA in conjunction with leading-twist transversity
GPDs, is very strong due to the large mass parameter $\mu_\pi$.
Twist-3 effects may be also generated by twist-3 GPDs 
\ci{Belitsky:2000vk}. 
However, for these GPDs there is no similar enhancement
known. Therefore, the contribution from the twist-3 GPDs is expected
to be small and neglected by us.

With the help of the relation between the 2-particle \da s, $\phiDA_p$ and $\phiDA_\sigma$,
and the 3-particle DA, $\phiDA_{3\pi}$, the twist-3 subprocess amplitude can solely be expressed by 
the latter \da. This manifestly demonstrates the vanishing of the 
twist-3 contribution in the Wandzura-Wilczek approximation which we observed 
previously \ci{signatures}. For the numerical analysis we thus have only to specify
the 3-particle \da. The parameters $f_{3\pi}$ and $\omega_{10}$ are taken
from the literature \ci{ball98}. A third parameter, $\omega_{20}$, is fitted to the
recent CLAS data on $\pi^0$ photoproduction at $s=11.06\,\gev^2$ \ci{clas-pi0}. The twist-2
subprocess amplitude is taken from our previous work \ci{signatures}, for the twist-2 \da{}
we used a recent lattice gauge theory results \ci{braun15}. The form factors, $R_V$ and $R_T$,
representing $1/x$-moments of GPDs, are taken from the GPD-analysis of the electromagnetic
form factors of the nucleon \ci{DK13} while for the form factor $R_A$ a result advocated for
in \ci{kroll17} is employed. The transversity form factors, $S_T$, $\bar{S}_T$ and $S_S$,
are evaluated from the transversity GPDs discussed in \ci{GK6} which describe fairly well 
exclusive electroproduction of pions at small $-t$. These GPDs are extrapolated to the large $-t$
region. 

Our results for the $\pi^0$ cross section agree rather well the recent CLAS data \ci{clas-pi0}.
It turns out that the twist-3 contribution dominates by far, the twist-2 contribution is almost 
negligible. Thus, we observed the same situation for wide-angle $\pi^0$ photoproduction  as
for deeply virtual $\pi^0$ electroproduction. We also presented predictions for the cross
section at other energies and for a number of spin-dependent observables. In particular
noteworthy are the helicity correlations $A_{LL}$ and $K_{LL}$. In contrast to wide-angle
Compton scattering where $A_{LL}= K_{LL}$ \ci{HKM} they are nearly mirror symmetric
(i.e.\ $A_{LL}\simeq -K_{LL}$) in wide-angle  photoproduction of $\pi^0$ or $\eta$ mesons.
This result is a consequence of the twist-3 dominance.  

The twist-3 mechanism we have proposed applies to the $s - t$
crossed process, $p\bar{p}\to \gamma\pi^0$, too. Also for
that process the twist-2 contribution falls short in comparison
with experiment \ci{KS05}. A hint at a dominant higher-twist
contribution to this process comes from the FERMI lab E760
experiment \ci{E760} which clearly have a stronger energy
dependence  than predicted by dimensional counting.

{\it Acknowledgements} We are grateful to Han Wen Huang who collaborated with us
at an early stage of the work. We also would like to thank 
Volodya Braun, Dimitry Ivanov,
Rainer Jakob, Dieter M\"{u}ller, Lech Szymanowski 
and Samuel Wallon for discussions and comments.
This work has been supported in part by the Croatian
Science Foundation (HrZZ) project ``Physics of Standard Model and
beyond'' HrZZ 5169, and the H2020 CSA Twinning project No. 692194,
''RBI-T-WINNING''.

\begin{appendix}
\renewcommand{\theequation}{\Alph{section}.\arabic{equation}}%
\setcounter{equation}{0}
\section{The 2- and 3-particle twist-3 pion \da s}
\label{sec:das}
In this section we supplement the main part of the paper 
by summarizing the definitions of the 2- and 3-particle pion 
distribution amplitudes up to twist-3 \ci{braun90,ball98,mannel} and 
the equations that relate them. 
For notational convenience we quote the vacuum-pion matrix elements 
for a charged pion. Their generalization to the case of a $\pi^0$ is obvious. 
One has to write the quark field operators as 
\be
\frac1{\sqrt{2}} \langle 0| \bar{u}\Gamma u -\bar{d}\Gamma d|\pi^0\rangle
\ee
All what we present in this appendix can straightforwardly be generalized to other pseudoscalar mesons.

The twist-2 \da{} is defined by the vacuum-pion matrix element
\be
\langle 0\mid \bar{u}(z_2) \gamma_\mu\gamma_5 \, d(z_1)\mid \pi^-(q')\rangle \=
      i f_\pi q'_\mu \int_0^1 d\tau \, e^{-i(\taub q'\cdot z_2+\tau q'\cdot z_1) } \,\phiDA_\pi(\tau)\,.
\label{eq:twist2}
\ee
The 2-particle twist-3 \da s are defined by
\ba
\langle 0\mid \bar{u}(z_2)\, \gamma_5 \, d(z_1)\mid \pi^-(q')\rangle &=&
      i f_\pi \mu_\pi \int_0^1 d\tau \, e^{-i(\taub q'\cdot z_2+\tau q'\cdot z_1) } \,\phiDA_{\pi p}(\tau)\,, 
\nn\\
\langle 0\mid \bar{u}(z_2) \sigma_{\mu \nu}\gamma_5  d(z_1)\mid \pi^-(q')\rangle &=&
    \frac{i}{6} f_\pi \mu_\pi \big(q'_\mu \, z_\nu - q'_\nu \, z_\mu\big) \nn\\
           &\times& \int_0^1 d\tau \, e^{-i(\taub q'\cdot z_2+\tau q'\cdot z_1) } \,\phiDA_{\pi\sigma}(\tau)\,.
\label{eq:2-particle-twist3}
\ea
Here $z=z_2-z_1$ ($z=[0,z^-, 0_\perp]$) and we take the massless limit, $q'{}^2=0$. We remind the reader 
that we are working in light-cone gauge. All Wilson lines, i.e. the path-ordered exponentials of 
the gluon fields,  are unity in that gauge. The local limits of the $\gamma_\mu\gamma_5$ and 
$\gamma_5$ matrix elements 
\ba
\langle 0\mid \bar{u}(0) \gamma_\mu\gamma_5 \, d(0)\mid \pi^-(q')\rangle &=&if_\pi q'_\mu\,, \nn\\
\langle 0\mid \bar{u}(0) \gamma_5 \, d(0)\mid \pi^-(q')\rangle &=& i f_\pi \mu_\pi\,,
\ea
provide constraints on the \da s $\phiDA_\pi$ and $\phiDA_{\pi p}$:
\be
\int_0^1 d\tau \phiDA_\pi(\tau) \= \int_0^1 d\tau \phiDA_{\pi p}(\tau)\=1
\label{eq:constraints}
\ee
as one sees with the help of translation invariance. There is no such constraint on $\phiDA_{\pi\sigma}$.
Its normalization is fixed by the equation of motion as we will see below.
The definitions \req{eq:twist2} and \req{eq:2-particle-twist3} can be combined into
\begin{eqnarray}
\langle 0\mid \bar{u}^g(z_2) d^f(z_1)\mid \pi^-(q')\rangle 
      &=& \frac{i f_\pi}{4} \,  
\frac{\delta_{fg}}{N_C}  \, \int_0^1 d\tau \, e^{-i(\taub q'\cdot z_2+\tau q'\cdot z_1) } \, \nn\\
     &\hspace*{-0.2\tw}\times& \hspace*{-0.1\tw}\left\{
      \gamma_5 \not \! q' \phiDA_\pi + \mu_{\pi} \gamma_5 \left[ \phiDA_{\pi p} 
            - \sigma_{\mu \nu} q'^{\mu} x^{\nu} \phiDA_{\pi\sigma} 
                     \right]\ \right\}
\,,
\label{eq:2-particle-matrixel}
\end{eqnarray}
leading to \req{eq:projector-2}.

The 3-particle twist-3 \da{} is defined by the quark-antiquark-gluon vacuum-pion matrix element
\ba
\lefteqn{
\langle 0|\bar{u}(z_b)\,\sigma_{\mu\nu}\gamma_5 \,g \,G_{\alpha\beta}(z_g) \, d(z_a)|\pi(q')\rangle } 
                 \hspace*{0.2\tw}\nn\\&= &  i f_{3\pi}\,\Big[q'_\alpha (q'_\mu g_{\nu\beta} 
          - q'_\nu g_{\mu\beta}) - (\alpha\leftrightarrow \beta)\Big] \nn\\
    & &    \times \int [d\tau]_3\; e^{-i(z_a\tau_a+z_b\tau_b+z_g\tau_g)} \phiDA_{3\pi}(\tau_a,\tau_b,\tau_g)
\label{eq:3-particle}
\ea
where
\be
[d\tau]_3=d\tau_a \, d\tau_b \, d\tau_g \, \delta(1-\tau_a-\tau_b-\tau_g)\,.
\ee
As usual also the 3-particle \da{} is normalized to
\be
\int [d\tau]_3  \, \phiDA_{3\pi}(\tau_a,\tau_b,\tau_g)\=1\,.
\label{eq:measure}
\ee

The 2-particle and 3-particle twist-3 \da s are connected through the equations of motion:
\ba
0 &=& \langle 0 \left| \bar{u}(z_2) \,i \slashed{D}_{z_1} \, d(z_1) \right| \pi^-(q') \rangle  \nn\\ 
&=&\langle 0 \left| \bar{u}(z_2) \,\left( i \slashed{\partial}_{z_1} \right)\, d(z_1) \right| 
                                                    \pi^-(q')\rangle \nn\\
  &+& \langle 0 \left| \bar{u}(z_2) \,\slashed{A}((z_1) \, d(z_1) \right| \pi^{-}(q') \rangle
\,.
\label{eq:eq-motion}
\ea
An analogous equation holds for the antiquark field. Using \req{eq:2-particle-matrixel} and
\req{eq:da-projector3}, derived for light-cone gauge, 
we cast the equations of motion into the simple closed form  
\be
\tau\phiDA_{\pi p}(\tau) + \frac16 \tau \phiDA_{\pi\sigma}'(\tau) - \frac13 \phiDA_{\pi\sigma}(\tau) \=  
                   \phiDA_1^{\rm EOM}(\tau) 
\label{eq:EOM1}
\ee
and for the antiquark
\be
\taub\phiDA_{\pi p}(\tau) - \frac16 \taub \phiDA_{\pi\sigma}'(\tau) - \frac13 \phiDA_{\pi\sigma}(\tau) \=  
                              \phiDA_2^{\rm EOM}(\tau) 
\label{eq:EOM2}
\ee
where
\be
\phiDA_1^{\rm EOM}(\tau)\=2\eta_3\int_0^{\tau} \frac{d\tau_{g}}{\tau_g}\,
                          \phiDA_{3\pi}(\tau-\tau_g,\taub,\tau_g)\,,
\label{eq:phi-eom1}
\ee
and
\be
\phiDA_2^{\rm EOM}(\tau)\=2\eta_3\int_0^{\taub} \frac{d\tau_{g}}{\tau_g}\,
                          \phiDA_{3\pi}(\tau,\taub-\tau_g,\tau_g)\,.
\label{eq:phi-eom2}
\ee
The prefactor $\eta_3$ is defined by
\be
\eta_3\= \frac{f_{3\pi}}{f_\pi\mu_\pi}\,.
\label{eq:eta3-def}
\ee
We stress that both the 2- and 3-particle pion \da s are symmetric under the exchange of the quark 
and antiquark momentum fractions, \req{eq:phi-eom1} and \req{eq:phi-eom2} are related by 
the replacement $\tau \leftrightarrow \taub$. Similar relations as \req{eq:EOM1} and \req{eq:EOM2} 
have been derived in the  light-cone gauge for the twist-3 \da s of a transversely polarized 
$\rho$ meson in \ci{wallon}.

A suitable combination of \req{eq:EOM1} and \req{eq:EOM2} leads to a first order linear differential
equation for the \da{} $\phiDA_\sigma$ which can easily be solved. The other \da{}, $\phiDA_p$, can 
subsequently be determined from  \req{eq:EOM1} for instance. 
We find for the 2-particle twist-3 \da s
\ba
\phiDA_{\pi\sigma}(\tau)&=&6 \tau \taub \left(\int d\tau
    \frac{\taub \phiDA_1^{\rm EOM}(\tau)-\tau \phi_2^{\rm EOM}(\tau)}{2 \tau^2 \taub^2} + C \right)\,, \nn\\
                    \phiDA_{\pi p}(\tau)&=& \frac{1}{6 \tau \taub} \phiDA_{\pi\sigma}(\tau) 
                 + \frac{1}{2\tau} \phiDA_{1}^{\rm EOM}(\tau)
                 + \frac{1}{2\taub} \phiDA_{2}^{\rm EOM}(\tau)\,.
\label{eq:2particle-det}
\ea
Using the 3-particle \da{} \req{eq:3-particle-da} and fixing the constant of integration, $C$, such 
that the constraint \req{eq:constraints} on $\phiDA_{\pi p}$ is respected, we find
\be
C\= \Big[1 + \eta_3 \big( 7\omega_{1,0} - 2\omega_{2,0} - \omega_{1,1}\big)\Big]
\ee
and 
\be
\phiDA_{\pi p} \= 1 +  \sum_{n=2,4,\ldots} a^p_{\pi n} C_n^{(1/2)}(2\tau-1)
\label{eq:phip}
\ee
with the Gegenbauer coefficients   ($a^p_{\pi n}=0$ for $n\geq 6$)
\be
a^p_{\pi 2}\=-\frac{10}{3} a^p_{\pi 4} \= \frac{10}{7} \eta_3\big(7\omega_{1,0} - 2\omega_{2,0} - \omega_{1,1}\big)\,.
\label{eq:P-gegenbauer}
\ee
Obviously, the second 2-particle twist-3 \da{} is not normalized to unity 
but we achieve that by a renormalization
\be
\phiDA_{\pi\sigma}\=\eta_\sigma \tilde{\phiDA}_{\pi\sigma}
\ee
with
\be
\tilde{\phiDA}_{\pi\sigma}\=6\tau\taub\Big[1 + \sum_{n=2,4,\ldots} a^\sigma_{\pi n} C_n^{(3/2)}(2\tau-1)\Big]
\label{eq:phis}
\ee
In this case the Gegenbauer coefficients read ($a^\sigma_{\pi n}=0$ for $n\geq 6$)
\ba
a^\sigma_{\pi 2} &=& \frac16\,\frac{\eta_3}{\eta_\sigma}\big(12 + 3\omega_{1,0} - 4\omega_{2,0}\big) \nn\\
a^\sigma_{\pi 4} &=& \frac1{105}\,\frac{\eta_3}{\eta_\sigma}\big(22\omega_{2,0} - 3\omega_{1,1}\big)
\label{eq:S-gegenbauer}
\ea
and with
\be
\eta_\sigma\=1 - \eta_3\big(12 - 4\omega_{1,0} + \frac87\omega_{2,0} + \frac47\omega_{1,1}\big)\,. 
\label{eq:S-norm}
\ee  
In the limit $\eta_3\to 0$, i.e.\ if the 3-particle \da{} is ignored, the 2-particle twist-3 \da s
reduce to the Wandzura-Wilczek approximation \req{eq:WW}
\be
\phiDA_{\pi p} \to \phiDA_{p}^{WW}\,,  \phiDA_{\pi\sigma}\to \phiDA_{\sigma}^{WW}\,, \eta_\sigma\to 1
\ee
In \ci{braun90,ball98,mannel,gorsky} the Fock-Schwinger gauge has been used instead of the 
light-cone one. With the Fock-Schwinger gauge one obtains a recursion formula for the moments of the 
various twist-3 pion \da s from which one can also determine 2-particle twist-3 \da s for a given 
3-particle \da.
However, the 2-particle \da s determined from the recursion formula  differ from our ones 
for the same 3-particle \da{} markedly. If the Fock-Schwinger gauge is employed the Wilson 
lines are not unity and are of significance.
We expect that a consistent calculation of the subprocess amplitudes using either the light-cone
gauge or the Fock-Schwinger one in the vacuum-particle matrix elements one leads to the same results. 
At least for the case of electroproduction of a transversely polarized $\rho$-meson  the equivalence 
of the two methods has been shown \ci{wallon}.

\section{2-particle and 3-particle twist-3 contributions to subprocess amplitudes}
\label{sec:tw3res}
\setcounter{equation}{0}

In this section we list and comment on the separate 2- and 3-particle twist-3 contributions. 
Their sum is given in \req{eq:twist3-sub}. As mentioned above, the 2-particle contribution
can be completely expressed through the 
combination of 2-particle twist-3 \da s appearing on the left-hand side 
of the equations of motion \req{eq:EOM1}\, \req{eq:EOM2} 
and can thus be simplified to
\ba
{\cal H}_{0-\lambda,\,\mu\lambda}^{twist-3,2-particle}
      &=&  4 \pi \als(\muR)\, f_\pi \,\mu_\pi \, \frac{C_F}{N_C} \,
          \frac{\sqrt{-\uh \sh}}{\sqrt{2}\sh^2\uh^2}
          \int_0^1 d\tau \, \phi^{\rm EOM}_{2} (\tau) \nn \\[0.2em]
 & \times& \left[ \left( \frac{2 \lambda-\mu }{2 (1-\tau)^2} \, + 
 \frac{2 \lambda+\mu }{2 \tau (1-\tau)} \, \right)\,(\sh^2+\uh^2) \right.  \nn \\[0.2em]
       & + & \left. \quad \mu\, \frac{\th\sh}{\tau (1-\tau)} \,\right]\, . 
\label{eq:tw3HeomSIMPLE}
\ea

On the other hand, for the 3-particle twist-3 amplitude we derive
\ba
{\cal H}_{0-\lambda,\,\mu\lambda}^{twist-3,3-particle} &=&  4 \pi \als(\muR) \, \frac{f_{3\pi}}{N_C} 
               \, \frac{\sqrt{-\uh \sh}}{\sqrt{2}\sh^2\uh^2} \, \nn \\[0.2em]
           & \times&  \int_0^1 d\tau \int_0^{1-\tau} \frac{d\tau_g}{\tau_g} 
             \, \phiDA_{3\pi}(\tau,1-\tau-\tau_g,\tau_g) \nn \\[0.2em]
            &\times& \left\{ \left(2 \lambda - \mu\right) \left[C_F \left(\frac{1}{1-\tau}
              - \frac{1}{1-\tau-\tau_g} \right) \frac{\sh^2+\uh^2}{\tau_g}\,\right.\right. \nn \\[0.2em]
    &+& \left. \left. \quad \left(C_F -\frac{C_A}{2} \right) \left(\frac{1}{\tau}+\frac{1}{1-\tau-\tau_g}
        \right) \frac{\th^2}{\tau_g} \, \right]\right. \nn \\[0.2em]
       &-& \left. \quad \frac{2\lambda + \mu }{\tau(1-\tau)}\left(\sh^2+\uh^2 \right)
              - 2\mu\,\frac{\sh\th}{\tau(1-\tau)} \right\}\, .
\label{eq:twist3-Hres3p}
\end{eqnarray}
The 2- and 3-particle twist-3 contributions do not respect current conservation separately,
only their sum do so. However, the terms proportional to the color factor $C_A$ occuring only
in \req{eq:twist3-Hres3p}, is gauge invariant separately.
Note that both \req{eq:tw3HeomSIMPLE} and \req{eq:twist3-Hres3p} possess terms that are not symmetric in
$\sh \leftrightarrow \uh$, i.e., do not obey the crossing properties expected 
for this process on general grounds \ci{CGLN}. However, their sum \req{eq:twist3-sub} is symmetric 
in $\sh \leftrightarrow \uh$, i.e., the expected crossing symmetry is recovered in the full twist-3 
contribution.

\end{appendix}


\begin{thebibliography}{99}

\bibitem{rad98} A.~V.~Radyushkin,
  Phys.\ Rev.\ D {\bf 58}, 114008 (1998)
  [hep-ph/9803316].

\bibitem{DFJK1} M.~Diehl, T.~Feldmann, R.~Jakob and P.~Kroll,
  Eur.\ Phys.\ J.\ C {\bf 8}, 409 (1999)
  [hep-ph/9811253].

 \bibitem{DK13} M.~Diehl and P.~Kroll,
  Eur.\ Phys.\ J.\ C {\bf 73}, no. 4, 2397 (2013)
  [arXiv:1302.4604 [hep-ph]].

\bibitem{hallA} A.~Danagoulian {\it et al.} [Hall A Collaboration],
  Phys.\ Rev.\ Lett.\  {\bf 98}, 152001 (2007)
  [nucl-ex/0701068 [NUCL-EX]].

\bibitem{huang00}  H. W. Huang and P. Kroll, 
                Eur.\ Phys.\ J.\  {\bf C17}, 423 (2000), [hep-ph/0005318].

\bibitem{signatures} H.~W.~Huang, R.~Jakob, P.~Kroll and K.~Passek-Kumeri\v{c}ki,
  Eur.\ Phys.\ J.\ C {\bf 33}, 91 (2004).
  [hep-ph/0309071].

\bibitem{ji-hoodbhoy} P.~Hoodbhoy and X.~D.~Ji,
  Phys.\ Rev.\ D {\bf 58}, 054006 (1998)
  [hep-ph/9801369].

\bibitem{diehl01} M.~Diehl,
  Eur.\ Phys.\ J.\ C {\bf 19}, 485 (2001)
  [hep-ph/0101335].

\bibitem{hermes08} A.~Airapetian {\it et al.} [HERMES Collaboration],
  Phys.\ Lett.\ B {\bf 682}, 345 (2010)
  [arXiv:0907.2596 [hep-ex]].


\bibitem{collins} J.~C.~Collins, L.~Frankfurt and M.~Strikman,
  Phys.\ Rev.\ D {\bf 56}, 2982 (1997)
  [hep-ph/9611433].

\bibitem{GK5} S.~V.~Goloskokov and P.~Kroll,
  Eur.\ Phys.\ J.\ C {\bf 65}, 137 (2010)
  [arXiv:0906.0460 [hep-ph]].

\bibitem{GK6} S.~V.~Goloskokov and P.~Kroll,
  Eur.\ Phys.\ J.\ A {\bf 47}, 112 (2011)
  [arXiv:1106.4897 [hep-ph]].

\bibitem{liuti}G.~R.~Goldstein, J.~O.~Gonzalez Hernandez and S.~Liuti,
  Phys.\ Rev.\ D {\bf 91}, no. 11, 114013 (2015)
  [arXiv:1311.0483 [hep-ph]].

\bibitem{defurne} M.~Defurne {\it et al.} [Jefferson Lab Hall A Collaboration],
  Phys.\ Rev.\ Lett.\  {\bf 117}, no. 26, 262001 (2016)
  [arXiv:1608.01003 [hep-ex]].

\bibitem{compass} A.~Sandacz [COMPASS collaboration],
 talk presented at the XVII workshop on High Energy Spin Physics,
Dubna, Russia, September 2017.

\bibitem{braun90} V.~M.~Braun and I.~E.~Filyanov,
  Z.\ Phys.\ C {\bf 48}, 239 (1990)
  [Sov.\ J.\ Nucl.\ Phys.\  {\bf 52}, 126 (1990)]
  [Yad.\ Fiz.\  {\bf 52}, 199 (1990)].

\bibitem{huang02} M.~Diehl, T.~Feldmann, H.~W.~Huang and P.~Kroll,
  Phys.\ Rev.\ D {\bf 67}, 037502 (2003)
  [hep-ph/0212138].


\bibitem{DFJK4} M.~Diehl, T.~Feldmann, R.~Jakob and P.~Kroll,
  Eur.\ Phys.\ J.\ C {\bf 39}, 1 (2005)
  [hep-ph/0408173].

\bibitem{kitagaki} T.~Kitagaki {\it et al.},
  Phys.\ Rev.\ D {\bf 28}, 436 (1983).

\bibitem{kroll17} P.~Kroll,
  Eur.\ Phys.\ J.\ A {\bf 53}, no. 6, 130 (2017)
  [arXiv:1703.05000 [hep-ph]].


\bibitem{abm11} S.~Alekhin, J.~Bl\"{u}mlein and S.~Moch,
  Phys.\ Rev.\ D {\bf 86}, 054009 (2012)
  [arXiv:1202.2281 [hep-ph]].

\bibitem{dssv09} D.~de Florian, R.~Sassot, M.~Stratmann and W.~Vogelsang,
  Phys.\ Rev.\ D {\bf 80}, 034030 (2009)
  [arXiv:0904.3821 [hep-ph]].

\bibitem{QCDSF06}M.~Gockeler {\it et al.} [QCDSF and UKQCD Collaborations],
  Phys.\ Rev.\ Lett.\  {\bf 98}, 222001 (2007)
  [hep-lat/0612032].

\bibitem{wallon} I.~V.~Anikin, D.~Y.~Ivanov, B.~Pire, L.~Szymanowski and S.~Wallon,
  Nucl.\ Phys.\ B {\bf 828}, 1 (2010)
  [arXiv:0909.4090 [hep-ph]].

\bibitem{anikin02} I.~V.~Anikin and O.~V.~Teryaev,
  Phys.\ Lett.\ B {\bf 554}, 51 (2003)
  [hep-ph/0211028].

\bibitem{beneke}M.~Beneke and T.~Feldmann,
  Nucl.\ Phys.\ B {\bf 592} (2001) 3
  [hep-ph/0008255].

\bibitem{two-gluon}  P.~Kroll and K.~Passek-Kumeri\v{c}ki,
  Phys.\ Rev.\ D {\bf 67}, 054017 (2003)
  [hep-ph/0210045].

\bibitem{PDG} C.~Patrignani {\it et al.} [Particle Data Group],
  Chin.\ Phys.\ C {\bf 40}, no. 10, 100001 (2016).

\bibitem{huang} T.~Huang, X.~H.~Wu and M.~Z.~Zhou,
  Phys.\ Rev.\ D {\bf 70}, 014013 (2004)
  [hep-ph/0402100].

\bibitem{ball98} P.~Ball,
  JHEP {\bf 9901}, 010 (1999)
  [hep-ph/9812375].

\bibitem{braun15} V.~M.~Braun, S.~Collins, M.~Göckeler, P.~Pérez-Rubio, A.~Sch\"{a}fer, 
  R.~W.~Schiel and A.~Sternbeck,
  Phys.\ Rev.\ D {\bf 92}, no. 1, 014504 (2015)
  [arXiv:1503.03656 [hep-lat]].

\bibitem{kogut-soper} J.~B.~Kogut and D.~E.~Soper,
  Phys.\ Rev.\ D {\bf 1}, 2901 (1970).

\bibitem{CGLN} G.~F.~Chew, M.~L.~Goldberger, F.~E.~Low and Y.~Nambu,
  Phys.\ Rev.\  {\bf 106}, 1345 (1957).

\bibitem{clas-pi0} M.~C.~Kunkel {\it et al.} [The CLAS Collaboration],
  arXiv:1712.10314 [hep-ex].

\bibitem{chernyak} V.~L.~Chernyak and A.~R.~Zhitnitsky,
  Phys.\ Rept.\  {\bf 112}, 173 (1984).

\bibitem{shi15} C.~Shi, C.~Chen, L.~Chang, C.~D.~Roberts, S.~M.~Schmidt and H.~S.~Zong,
  Phys.\ Rev.\ D {\bf 92}, 014035 (2015)
  [arXiv:1504.00689 [nucl-th]].

\bibitem{choi17}  H.~M.~Choi and C.~R.~Ji,
  Phys.\ Rev.\ D {\bf 95}, no. 5, 056002 (2017)
  [arXiv:1701.02402 [hep-ph]].

\bibitem{nam06}S.~i.~Nam and H.~C.~Kim,
  Phys.\ Rev.\ D {\bf 74}, 096007 (2006)
  [hep-ph/0608018].


\bibitem{HKM} H.~W.~Huang, P.~Kroll and T.~Morii,
  Eur.\ Phys.\ J.\ C {\bf 23}, 301 (2002)
  Erratum: [Eur.\ Phys.\ J.\ C {\bf 31}, 279 (2003)]
  [hep-ph/0110208].

\bibitem{glueX-sigma} H.~Al Ghoul {\it et al.} [GlueX Collaboration],
  Phys.\ Rev.\ C {\bf 95}, no. 4, 042201 (2017)
  [arXiv:1701.08123 [nucl-ex]].

\bibitem{fanelli} C.~Fanelli {\it et al.},
  Phys.\ Rev.\ Lett.\  {\bf 115}, no. 15, 152001 (2015)
  doi:10.1103/PhysRevLett.115.152001
  [arXiv:1506.04045 [nucl-ex]].

\bibitem{hamilton}D.~J.~Hamilton {\it et al.} [Jefferson Lab Hall A Collaboration],
  Phys.\ Rev.\ Lett.\  {\bf 94}, 242001 (2005)
  [nucl-ex/0410001].

\bibitem{bogdan}  B.~Wojtsekhowski in S.~Ali {\it et al.},
  arXiv:1704.00816 [nucl-ex].

\bibitem{FKS1} T.~Feldmann, P.~Kroll and B.~Stech,
  Phys.\ Rev.\ D {\bf 58}, 114006 (1998)
  [hep-ph/9802409].

\bibitem{frankfurt} L.~L.~Frankfurt, P.~V.~Pobylitsa, M.~V.~Polyakov and M.~Strikman,
  Phys.\ Rev.\ D {\bf 60}, 014010 (1999)
  [hep-ph/9901429].

\bibitem{ball06} P.~Ball, V.~M.~Braun and A.~Lenz,
  JHEP {\bf 0605}, 004 (2006)
  [hep-ph/0603063].

\bibitem{Belitsky:2000vk}
  A.~V.~Belitsky, D.~Mueller, A.~Kirchner and A.~Schafer,
  Phys.\ Rev.\ D {\bf 64} (2001) 116002
  [hep-ph/0011314].

\bibitem{KS05} 
P.~Kroll and A.~Sch\"{a}fer,
  Eur.\ Phys.\ J.\ A {\bf 26} (2005) 89
  [hep-ph/0505258].

\bibitem{E760} 
 T.~A.~Armstrong {\it et al.} [Fermilab E760 Collaboration],
  Phys.\ Rev.\ D {\bf 56} (1997) 2509.

\bibitem{mannel}A.~Khodjamirian, T.~Mannel and P.~Urban,
  Phys.\ Rev.\ D {\bf 67}, 054027 (2003)
  [hep-ph/0210378].

\bibitem{gorsky} A.~S.~Gorsky,
  ITEP-87-85.


\end{thebibliography}
\end{document}